\documentclass[manuscript]{aastex}

\usepackage{graphics,color}
\definecolor{red}{rgb}{1,0,0}

\shorttitle{Asteroid Rotation Periods From PTF}
\shortauthors{Chang et al.}
\begin{document}
%\begin{CJK*}{UTF8}{gbsn}
\title{313 new asteroid rotation periods from Palomar Transient Factory observations}

\author{Chan-Kao Chang\altaffilmark{1}; Wing-Huen Ip\altaffilmark{1,2}; Hsing-Wen Lin\altaffilmark{1};
Yu-Chi Cheng\altaffilmark{1}; Chow-Choong Ngeow\altaffilmark{1}; Ting-Chang Yang\altaffilmark{1}; Adam
Waszczak\altaffilmark{3}; Shrinivas R. Kulkarni\altaffilmark{4}; David Levitan\altaffilmark{4};
Branimir Sesar\altaffilmark{4}; Russ Laher\altaffilmark{5}; Jason Surace\altaffilmark{5}; Thomas. A.
Prince\altaffilmark{4} and the PTF Team}

\altaffiltext{1}{Institute of Astronomy, National Central University, Jhongli,Taiwan}
\altaffiltext{2}{Space Science Institute, Macau University of Science and Technology, Macau}
\altaffiltext{3}{Division of Geological and Planetary Sciences, California Institute of Technology,
Pasadena, CA 91125, USA} \altaffiltext{4}{Division of Physics, Mathematics and Astronomy, California
Institute of Technology, Pasadena, CA 91125, USA} \altaffiltext{5}{Spitzer Science Center, California
Institute of Technology, M/S 314-6, Pasadena, CA 91125, USA}

\email{rex@astro.ncu.edu.tw}

%\end{CJK*}

\begin{abstract}
A new asteroid rotation period survey have been carried out by using the {\it Palomar Transient
Factory} (PTF). Twelve consecutive PTF fields, which covered an area of 87 deg$^2$ in the ecliptic
plane, were observed in $R$ band with a cadence of $\sim$20 min during February 15--18, 2013. We
detected 2500 known asteroids with a diameter range of 0.5 km $\leq D \leq$ 200 km. Of these, 313
objects had highly reliable rotation periods and exhibited the ``spin barrier'' at $\sim2$ hours. In
contrast to the flat spin rate distribution of the asteroids with 3 km $\leq D \leq$ 15 km shown by
\citet{Pravec2008}, our results deviated somewhat from a Maxwellian distribution and showed a decrease
at the spin rate greater than 5 rev/day. One super-fast-rotator candidate and two possible binary
asteroids were also found in this work.
\end{abstract}


\section{Introduction}
Time-series photometry is a powerful tool to derive physical properties of solar system objects
including the rotation periods and shapes of asteroids and cometary nuclei. \citet{Harris1996} showed
a ``spin barrier'' at 2.2 hours for asteroids with $D \gtrsim 1$ km, which indicates large asteroids
are gravitationally bounded aggregations (i.e., rubble-pile structure). Following that study,
\citet{Pravec2000} revealed that the asteroids with diameter larger than a few hundred meters are
rubble-piles and have spin rates lower than the ``spin barrier'', while smaller asteroids may rotate
faster than the ``spin barrier'' \citep[i.e., super-fast-rotator; see an example study
of][]{Hergenrother2011} and are likely monolithic objects. Exceptions to this rule are rare. For
example, 2001 OE84 has a diameter of 0.9\,km and a rotation period of 29.19\,min \citep{Pravec2002}.
Subsequently, \citet{Holsapple2007} suggested a size-dependent strength for asteroids and predicted
the existence of kilometer size super-fast-rotators. Moreover, the spin rate distributions of
different size asteroids are useful to investigate how different mechanisms alter the asteroid
rotations. \citet{Salo1987} showed that a collisionally evolved asteroidal system should have a
Maxwellian spin rate distribution. This is true for asteroids with $D > 40$ km \citep{Pravec2002}.
However, mechanisms in addition to collision, in particular the Yarkovsky-O'Keefe-Radzievskii-Paddack
\citep[YORP;][]{Rubincam2000}, create excesses both at the very slow and fast ends for smaller size
asteroids \citep{Pravec2002, Polishook2008, Masiero2009}.

While this information is of importance in establishing the global property of the asteroids, only a
small fraction ($\sim$4000 among $\sim$620000 known asteroids) have published lightcurves and rotation
periods \citep{Warner2009}. In recent years large field of view surveys have been used to study
asteroid rotation periods \citep[see examples of][]{Masiero2009, Polishook2012}. With the increasing
volume of such data sets, the studies of the asteroid rotation properties (i.e., spin rate limit and
spin rate distribution) can be done as a function of different taxonomic types, dynamical groups and
asteroid families.

As demonstrated by the pilot work of \citet{Polishook2012}, the asteroidal observations in the {\it
Palomar Transient Factory} (PTF) synoptic survey could make important contributions in this respect.
They reported their analysis of four overlapping PTF fields covering 21 \,deg$^2$ with multiple
observations ($\geq 10$ images) per night and a typical cadence of $\sim20$ min. Of the 624 asteroids
detected in their work, 88 of them have well-determined rotation periods and 85 have low quality
rotation periods. Here, we continue this line of work by presenting 312 new, high quality, asteroid
rotation periods. The final goal of this long term project is to collect a sample of about
$\sim10^{4}$ asteroid rotation periods.

In Section 2, the observation information is described. The method of data analysis is given in
Section 3. The results on the statistical distribution of the rotation periods and the properties of
some individual asteroids are presented in Section 4. A summary and conclusion can be found in Section
5.

\section{Observations}
The PTF\footnote{http://ptf.caltech.edu} is a synoptic survey designed to explore the transient and
variable sky \citep{Law2009, Rau2009}, which employs the 48-inch Oschin Schmidt Telescope equipped
with a 11 chips mosaic CCD camera (i.e., the former CFHT-12K camera, in which the chip no. 3 is out of
function). The available filters include Mould-{\it R}, Gunn-{\it g\arcmin} and $H_{\alpha}$. Such a
configuration has a field of view of $\sim7.26$ \,deg$^2$ and a pixel scale of 1.01\arcsec. The
$5\sigma$ median limiting magnitude of an exposure of 60 s in $R$ band is $\sim$21 mag
\citep{Law2010}.

\begin{figure}
  %%\epsscale{0.8}
  \plotone{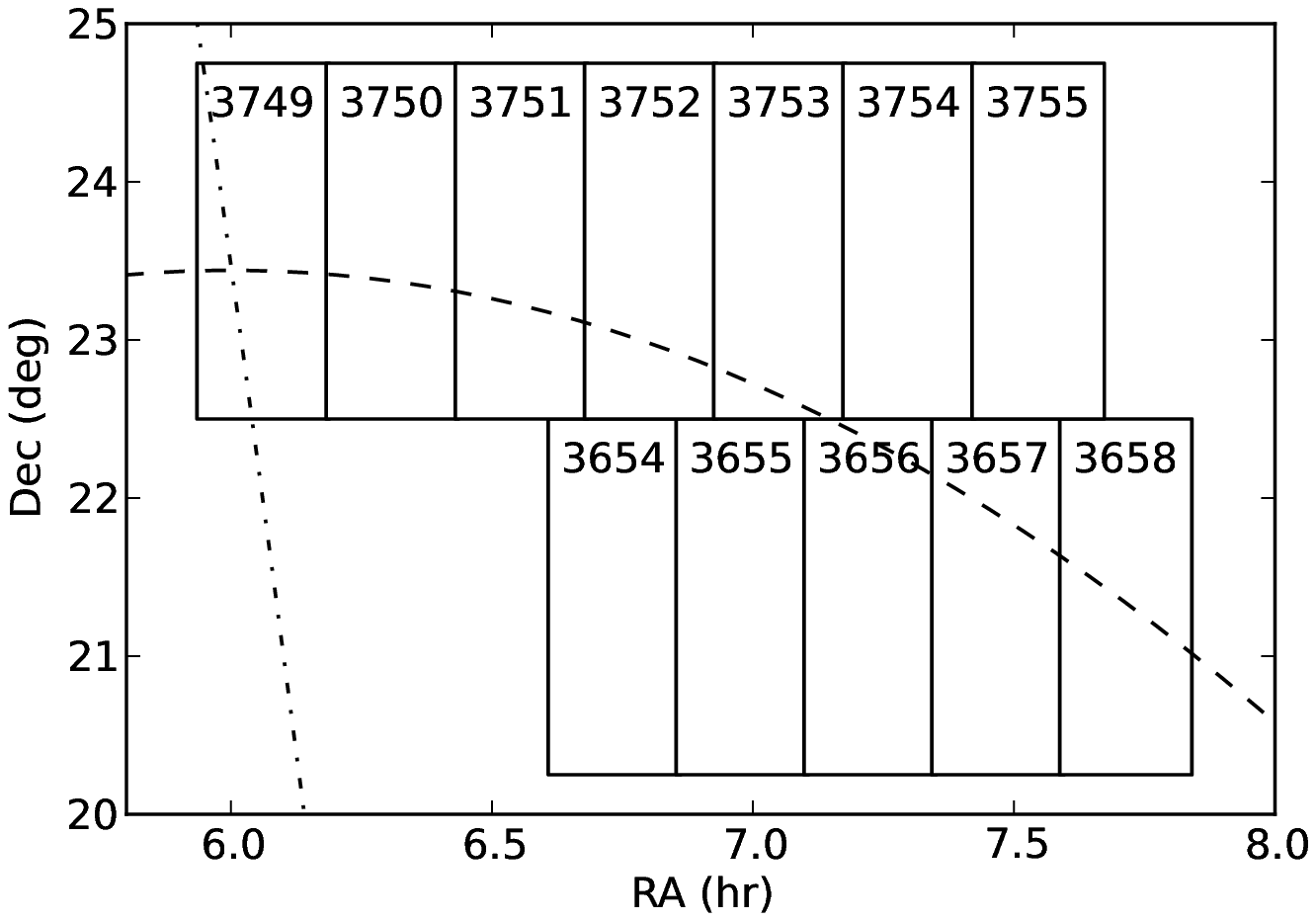}
  \caption{The configuration of 12 PTF fields. Each rectangular represents a PTF field with an field ID on top.
  The field of view of each PTF field is $3.50^\circ \times 2.31^\circ \sim 7.26$ \,deg$^2$. The dashed
  and dot-dashed lines show the ecliptic and Galactic planes, respectively.
  Note that the scales of R.A. and Declination are not in proper ratio.}
  \label{obs_fig}
  \end{figure}

As part of the Ten Thousand Asteroid Rotation Periods project (10kARPs), twelve consecutive PTF
fields, which covered an area of 87 \,deg$^2$ on the ecliptic plane, were observed in $R$ band with a
cadence of $\sim20$ min during February 15--18, 2013. The exposure time of each image was 60 s.
Fig.~\ref{obs_fig} shows the field configuration and Table~\ref{obs_log} lists the observation
information. Since this campaign was dedicated to the study of Galactic variables as well, our target
fields were close to the Galactic plane.

\section{Data Analysis}\label{sec_data}
\subsection{Data Reduction and Photometry Calibration}\label{data_cali}
Each PTF exposure was processed by the PTF photometric pipeline which included image splitting,
de-biasing, flat-fielding, generation of mask images, source extraction, astrometric calibration and
photometric calibration \citep[][Laher et al. 2014, PASP, submitted]{Grillmair2010}. The final
products of this pipeline included reduced images, mask images and source catalogs. The absolute
photometric calibration, described in \citet{Ofek2012a, Ofek2012b}, was done by using SDSS stars
\citep{York2000} and routinely reached a precision of $\sim0.02$ \,mag. In this work, we used source
catalogs computed by SEXtractor \citep{Bertin1996} to extract asteroid lightcurves and employed
relative (lightcurve-calibrated) photometry \citep[for algorithm details see][]{Levitan2011,
Ofek2012a} which typically had a relative photometry accuracy of $\sim3$ mmag and $\sim0.1$ mag in the
bright ($\sim15$ mag) and faint (i.e., $\sim19$ mag) ends, respectively \citep[][]{Agueros2011,
Levitan2011, Ofek2011}. The photometric calibration described above was done on night-, field- and
CCD-bases. Therefore, a systemic offset was introduced to each data set obtained from different CCDs,
fields and nights. This small offset will be corrected in the period fitting process, later described
in Section~\ref{period_analysis}.

\subsection{Light Curve Extraction of Known Asteroids}
%%We first remove the detections belonging to the stationary sources from the PTF source catalogs. To do
%%this, we construct a reference image for each PTF field/CCD based on about 10 images with the best
%%image quality (i.e., the best seeing values and the lowest backgrounds) and use SExtracor
%%\citep{Bertin1996} to detect the sources in the reference images to create reference catalogs. Then,
%%we perform a spatial cross match of 2\arcsec~ radius on PTF source catalogs against the reference
%%source catalogs to identify those detections with a stationary source counterpart and
%%
%%
%%Before extracting asteroid lightcurves, we remove the detections belonging to the stationary sources
%%from the PTF source catalogs. To do this, we construct reference images for each PTF field/CCD based
%%on about 10 images with the best image quality (i.e., the best seeing values and the lowest
%%backgrounds) and use SExtracor \citep{Bertin1996} to detect the sources in the reference images to
%%create reference catalogs. Then, we perform a spatial cross match of 2\arcsec~ radius on PTF source
%%catalogs against the reference source catalogs to let those detections without a stationary source
%%counterpart go through the comparison with asteroidal ephemerides which are .

The detections in a PTF source catalog were divided into stationary sources (i.e., the source would be
detected at the same position repeatedly) and non-stationary sources (e.g., moving objects and false
detections).
%%The detections in a PTF source catalog are from static (i.e., non-moving) sources and moving objects.
%%Only the detections without a static source counterpart were used in our lightcurve extraction.
To rule out the detections of stationary sources in the following asteroid lightcurve extraction, we
performed a spatial cross match with a radius of 1\arcsec~ on each source catalog against the
reference source catalog. To build the reference source catalog, we used three PTF source catalogs
with the best seeing of each night to pick out the sources that had been detected more than three
times in the same position (i.e., within 1\arcsec~ radius). The mean position of the detections of the
same reference source was assigned to be its RA and Dec. Then, we performed another spatial cross
match with a radius of 1\arcsec~ on the detections of non-stationary sources of each source catalog
against the ephemerides of the asteroids with $V \leq 22$ mag. The asteroidal ephemerides were
obtained from MPChecker\footnote{the online ephemerides service hosted at the Minor Planet Center;
http://scully.cfa.harvard.edu/cgi-bin/checkmp.cgi} according to our exposures. In the last, the
detections from the same asteroid were combined to generate its lightcurve. When a lightcurve
contained more than five detections, it was identified as a real event (hereafter, the PTF detected
asteroids).

\subsection{Photometric Stability Evaluation}
\begin{figure}
  %%\epsscale{0.8}
  \plotone{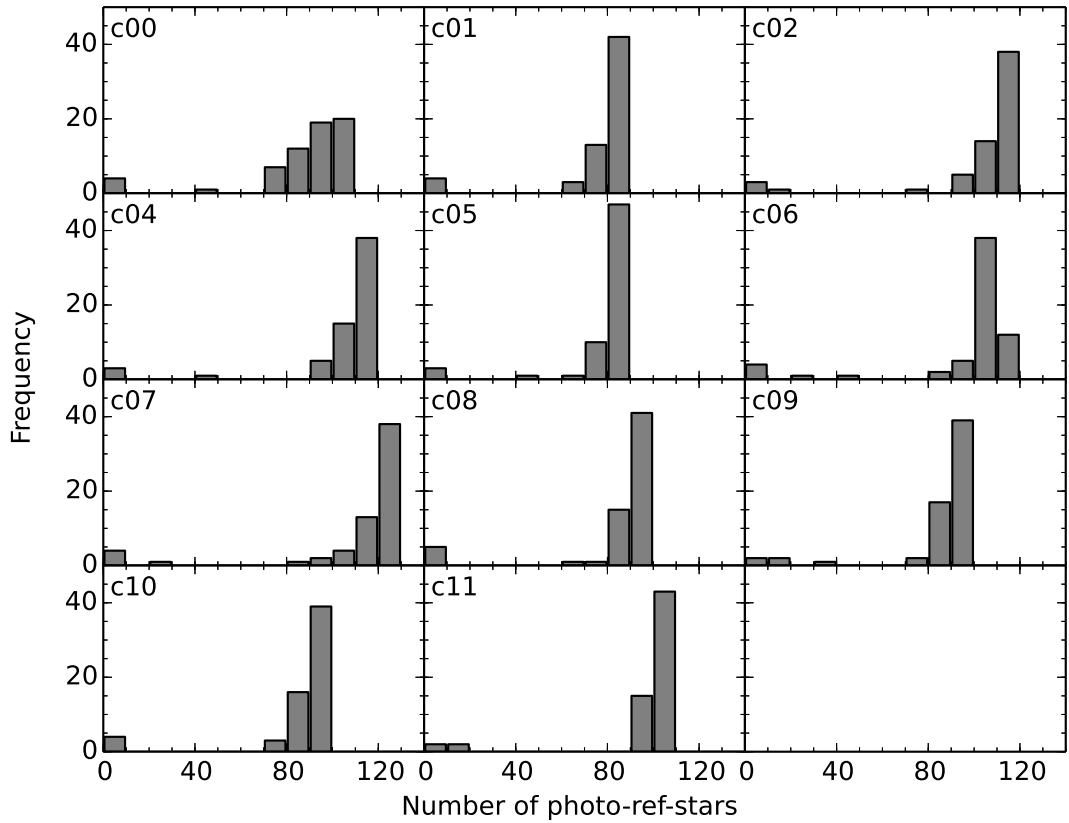}
  \caption{The number distribution of photo-ref-stars of 17.5 $<$ R $<$ 17.6 mag for each
  CCD in field 3655. The CCD numbers are shown on the upper-left corner of each plot.}
  \label{n_pho_ref}
  \end{figure}

The photometric stability of each source catalog was also evaluated. To do this, we chose the
reference sources in the 17--18 $R$ magnitude range that had standard deviations $\leq$ 0.075 mag
during the whole campaign (i.e., relatively brighter sources without brightness variations) to be the
photometric reference stars (hereafter, photo-ref-stars). Then, we grouped the photo-ref-stars into
bins of 0.1 mag and calculate their mean magnitudes and standard deviations for each source catalog.
To judge the photometric stability of each source catalog, we only considered those bins that had more
than 30 photo-ref-stars. When a source catalog had one of the following conditions, it would be
identified as photometrically unstable and not to be used in the asteroid rotation period analysis: a)
three or more bins had mean magnitudes out of the bin boundaries, b) three or more bins had standard
deviations $\geq$ 0.075 mag and c) six or more bins had less than 30 photo-ref-stars. In general, most
source catalogs (i.e., $\sim90$ \%) could fulfill the requirement described above. The rest failing to
pass the criteria were mostly due to the bad weather exposures which usually made the three conditions
happen at the same time. Fig.~\ref{n_pho_ref} shows the number distributions of the photo-ref-stars of
17.5 $<$ mag $<$ 17.6 of all source catalogs for each CCD in field 3655. Most catalogs had more than
80 photo-ref-stars and only several catalogs had less than 30. We also excluded the detections that
were flagged by the PTF photometric pipeline as artifacts (e.g., aircraft/satellite track, high dark
current pixel, noisy/hot pixel, saturated pixel, dead/bad pixel, ghost image, dirt on the optics,
CCD-bleed or bright star halo and the defects flagged by the SExtractor).

\subsection{Rotation Period Analysis}\label{period_analysis}
For measuring the synodic rotation periods of the PTF detected asteroids, the observing times were
corrected for light-travel time (i.e., the time interval of photon traveling from object to Earth) and
the magnitudes were reduced to both heliocentric ($r$) and geocentric $\triangle$ distances at 1 AU by
\begin{equation}\label{reduced_mag}
  M_{R (r=1,\triangle=1)} = R + 5\log(r\triangle),
\end{equation}
where $M_R$ is the $R$ band reduced magnitude. The orbital elements were obtained from the Minor
Planet Center\footnote{http://minorplanetcenter.net} and the heliocentric and geocentric distances
were calculated by the PyEphem\footnote{http://rhodesmill.org/pyephem/}.

Since the phase angles ($\alpha$) only had a small change in a campaign over four-nights, we applied
the $H$--$G$ system with a fixed $G_R$ slope of 0.15  to estimate the absolute magnitude $H_R$
\citep{Bowell1989}:
\begin{equation}\label{Hmag}
  H_R = \langle M_{R (r=1,\triangle=1)}\rangle+2.5\log[(1-G_R)\phi_1+G_R\phi_2],
\end{equation}
where $\phi_1$ and $\phi_2$ are the phase angle parameters as below:
\begin{equation}
  \phi_1 = \exp [-3.33 \tan(0.5\langle \alpha \rangle)^{0.63}],
\end{equation}
\begin{equation}
  \phi_2 = \exp [-1.87 \tan(0.5\langle \alpha \rangle)^{1.22}].
\end{equation}

\begin{figure}
  %%\epsscale{0.8}
  \plotone{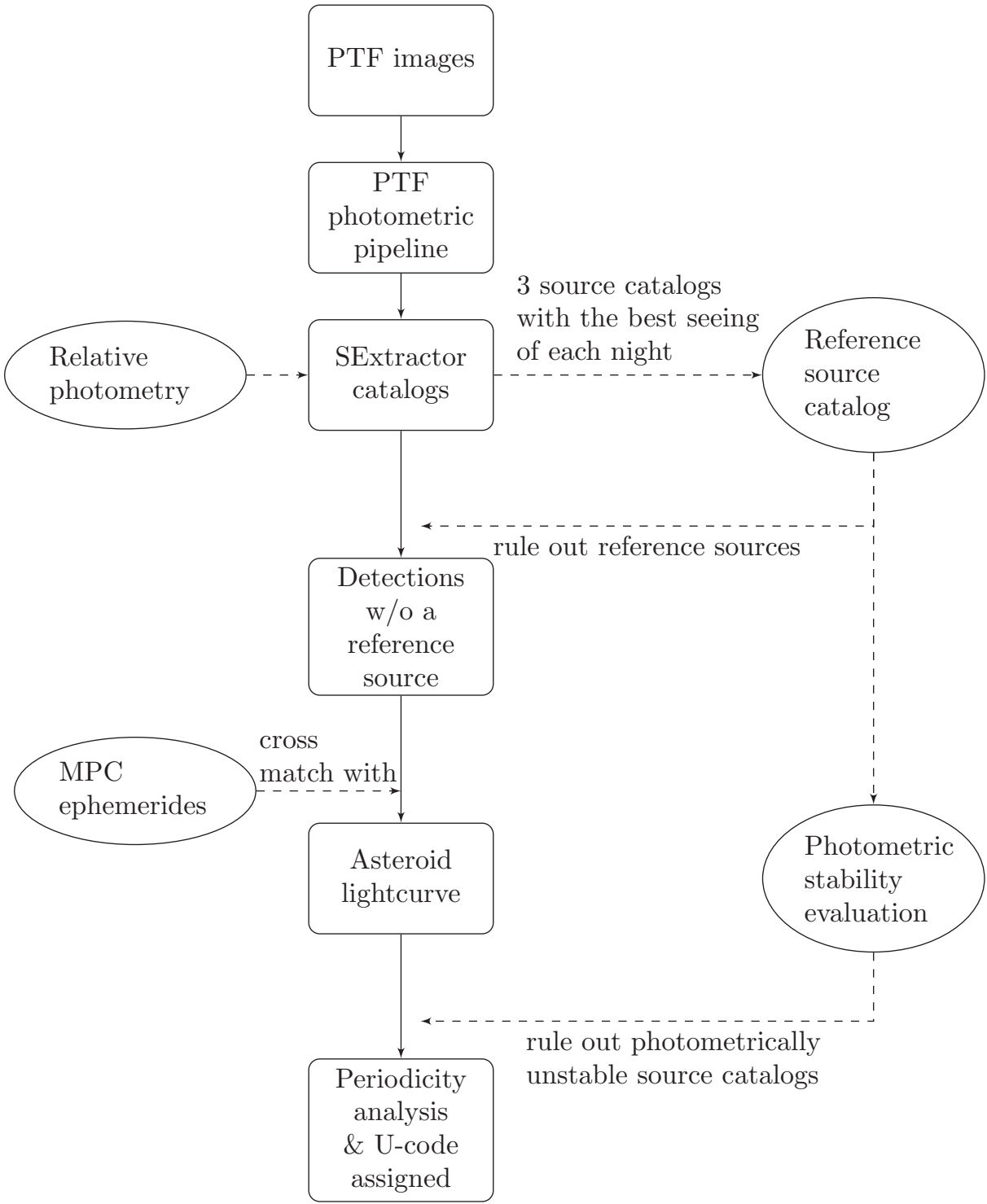}
  \caption{The data process flow chart. See section~\ref{sec_data} for explanation.}
  \label{flowchart}
  \end{figure}

Then, we fitted a second-order Fourier series to each asteroid lightcurve that had more than eight
detections to search the periodicity:
\begin{equation}\label{FTeq}
  M_{i,j} = \sum_{k=1,2}^{N_k} B_k\sin\left[\frac{2\pi k}{P} (t_j-t_0)\right] + C_k\cos\left[\frac{2\pi k}{P} (t_j-t_0)\right] + Z_i,
\end{equation}
where $M_{i,j}$ is the $R$ band reduced magnitude measured at the light-travel time corrected epoch
$t_j$, $B_k$ and $C_k$ are the Fourier coefficients, $P$ is the rotation period and $t_0$ is an
arbitrary epoch. As described in Section~\ref{data_cali}, the photometric calibration was carried out
on night-, field- and CCD-bases. Thus, we also fitted a constant value $Z_i$ in Eq. (\ref{FTeq}) to
correct the small systematic offsets between different data sets, where a data set was defined as all
the measurements taken on the same night, field and CCD with the subscript $i$ marking the $i$th data
set. The Eq.~(\ref{FTeq}) was solved by using least-squares minimization for each given $P$ to obtain
the other free parameters. We tried the 0.25--50 frequency range with a step of 0.0025 to cover the
majority of asteroid rotation periods \citep[e.g. about 20 min to about 80 h;][]{Pravec2000}. Then, we
reviewed all possible rotation periods (i.e., the periods with outstanding $\chi^2$ values from the
others) for each objects by inspecting the folded lightcurves to pick out the best result and assigned
a quality code $U$ as introduced by \citet[][; where `3' means highly reliable, `2' means some
ambiguity, `1' means possible but may be wrong and `0' means no detection]{Warner2009}. To estimate
the uncertainty of the derived rotation period, we calculated the range of periods with $\chi^2$
smaller than $\chi_{best}^2+\triangle\chi^2$, where $\chi_{best}^2$ is the $\chi^2$ of the pick out
period and $\triangle\chi^2$ is calculated from the inverse $\chi^2$ distribution assuming $1 + 2N_k +
N_s$ degrees of freedom. The amplitudes of the objects with full lightcurve coverage were adopted from
the second-order Fourier series fitting. However, this would probably underestimate the amplitudes of
the lightcurves of sharp minimum/maximum. Some folded lightcurves just covered part of a rotation
period and some only showed a single minimum due to their sparse data points, however, we still
assigned them $U = 2$ for their clear folded lightcurves. To give lower limits on the amplitude for
these objects, we calculated a 90\% magnitude range centered on the range median of their small offset
corrected lightcurves. This can reject the upper and lower 5\% detections to avoid those outliers
(i.e., the detections obviously deviated from the expected maximum/minimum of the folded lightcurve).
Such outliers could be the detections contaminated by the nearby bright stars or the artifacts not
filtered out from the lightcurve extraction. These objects need followup observations to confirm their
rotation periods. We will have more discussion of these cases in Section~\ref{discuss_p}. To have an
overview on our data analysis procedure, we show the flow chart in Fig.~\ref{flowchart}.

\section{Results}
\subsection{Detected asteroids}
\begin{figure}
  %%\epsscale{0.8}
  \plotone{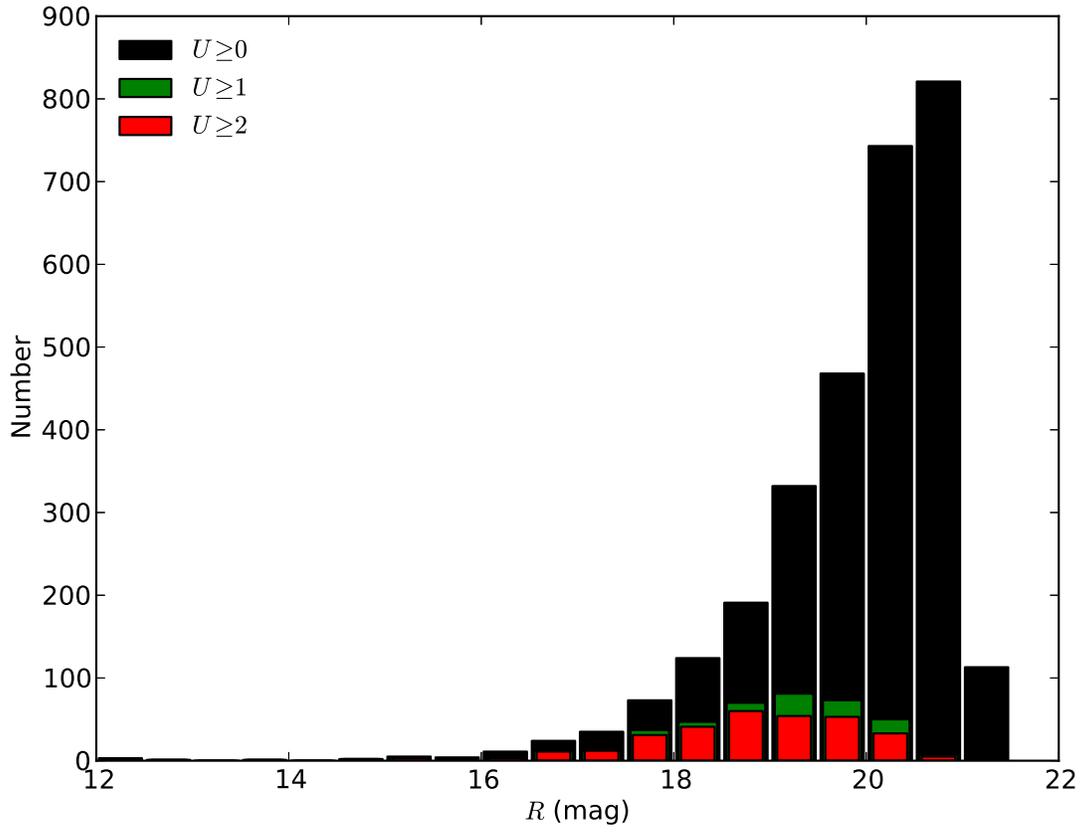}
  \caption{The distributions of magnitudes of the PTF detected asteroids. Black, green and red
  represent rotation periods of $U \geq 0$, $U \geq 1$ and $U \geq 2$, respectively.}
  \label{his}
  \end{figure}

\begin{figure}
  %%\epsscale{0.8}
  \plotone{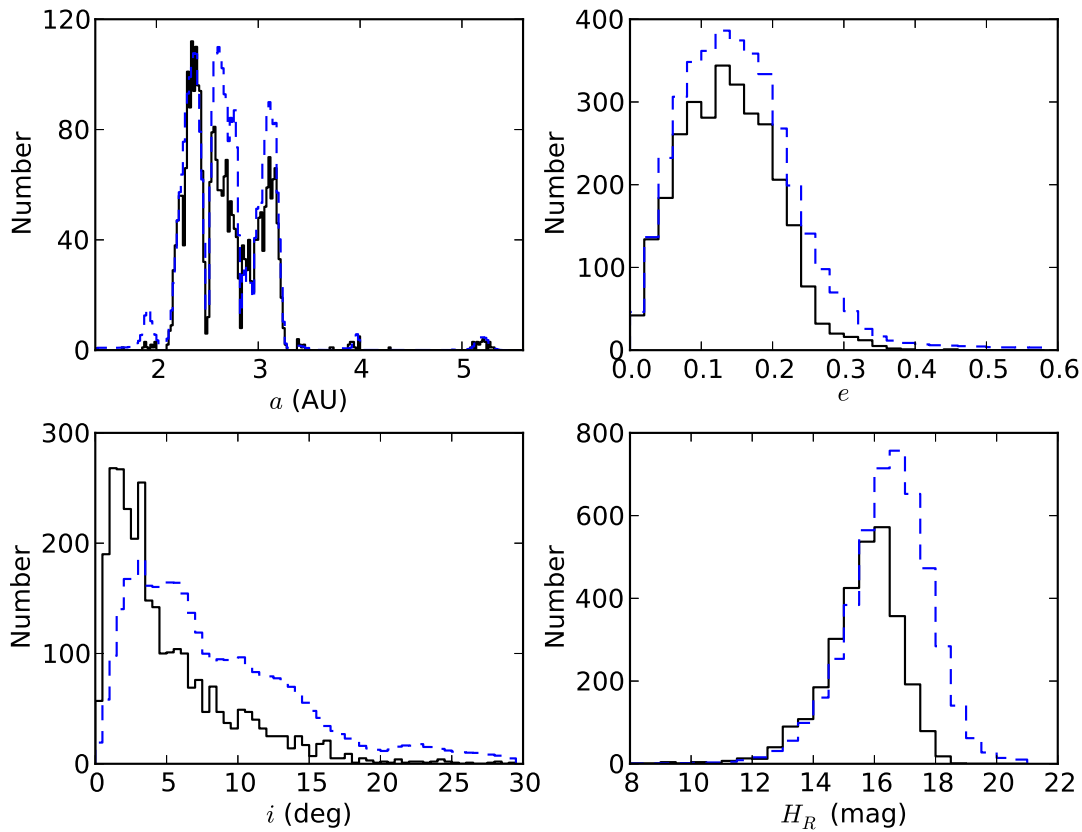}
  \caption{The distributions of orbital elements for the PTF detected asteroids (solid line)
  along with that of known asteroids with $a < 6$ AU in arbitrary normalization (dashed line).
  From the upper-left to the lower-right panels are the distributions for semimajor axes ($a$, 0.02 AU bins),
  eccentricities ($e$, 0.01 bins), inclinations ($i$, $0.5^\circ$ bins) and absolute magnitudes ($H_R$, 0.3 mag bins).}
  \label{aeih_dist}
  \end{figure}

There were 2500 PTF detected asteroids in a magnitude range of 14--22 mag in this work (see
Fig.~\ref{his}) and their distributions of the semimajor axes ($a$), eccentricities ($e$),
inclinations ($i$) and absolute magnitudes ($H_R$) along with that of all known asteroids with $a < 6$
AU are shown in Fig.~\ref{aeih_dist}. The majority of the PTF detected asteroids were main belt
asteroids and the others were a few Hilda, Jovian Trojan asteroids and near Earth objects. The PTF
detection rate is higher for the inner main belt asteroids and fair to all eccentricities. Since we
focused on the ecliptic plane, most PTF detected asteroids concentrate on low inclinations.

\begin{figure}
  %%\epsscale{0.8}
  \plotone{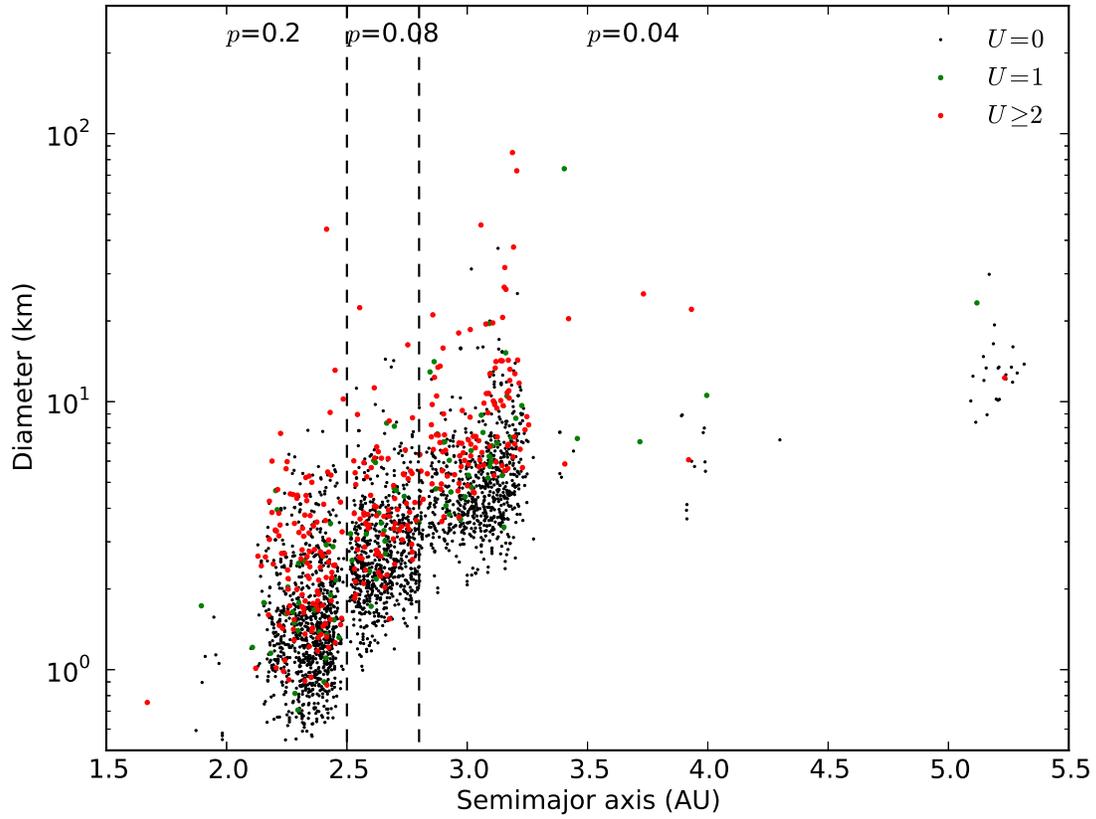}
  \caption{The plot of the diameter vs. semimajor axis for the PTF detected objects. The red, gray
  and black filled circles represent rotation periods of $U = 0$, $U = 1$ and $U \leq 2$, respectively. The dashed
  lines show the semimajor axis ranges for different empirical values of geometric albedo ($p$).}
  \label{a_d}
  \end{figure}

The diameters of the PTF detected asteroids were obtained from the preliminary results of
$WISE$/NEOWISE \citep{Grav2011, Mainzer2011, Masiero2011}. While the $WISE$/NEOWISE diameters were not
yet available, the diameters were estimated by the following equation
\begin{equation}\label{dia_eq}
  %%D = {1329 \over \sqrt{p_R}} 10^{-H_R/5}
  D = {1130 \over \sqrt{p_R}} 10^{-H_R/5},
\end{equation}
where $D$ is the diameter in units of km, $p_R$ is the geometric albedo in $R$ band and the conversion
constant, 1130, is adopted from \citet{Jweitt2013}. According to the semimajor axis ranges, we used
three empirical values for geometric albedo, which are (a) $p = 0.20$ for $a \le 2.5$ AU, (b) $p =
0.08$ for $2.5 < a \le 2.8$ AU and (c) $p = 0.04$ for $a> 2.8$ AU \citep{Tedesco2005}. The plot of the
semimajor axis vs. diameter for the PTF detected asteroids is given in Fig.~\ref{a_d}, which shows the
PTF was able to detect a few hundred meters size objects in inner main belt, kilometer size objects in
the outer main belt and 10 kilometer size objects for Hilda/Jovian Trojan asteroids.

\subsection{Derived Rotation Periods}\label{discuss_p}
\begin{figure}
  %%\epsscale{0.8}
  \plotone{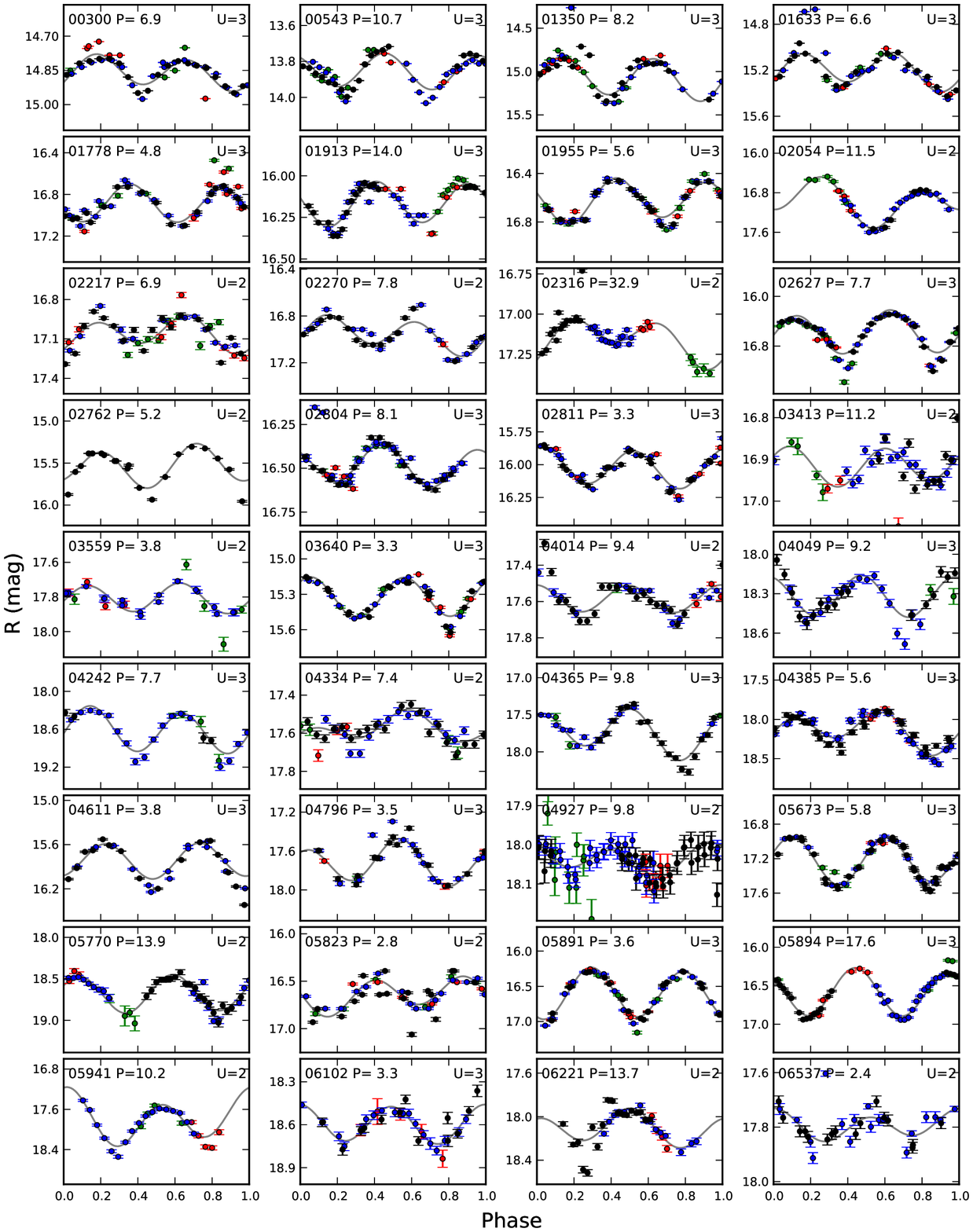}
  \caption{Set of 36 folded lightcurves for the PTF-U2 asteroids. The green, red, blue and black
  circles represent the observation date of February 15, 16, 17 and 18, 2013, respectively. The designation,
  rotation period in hours ($P$) and the quality code $U$ of each object are shown on each plot.}
  \label{lightcurve00}
  \end{figure}

\begin{figure}
  %%\epsscale{0.8}
  \plotone{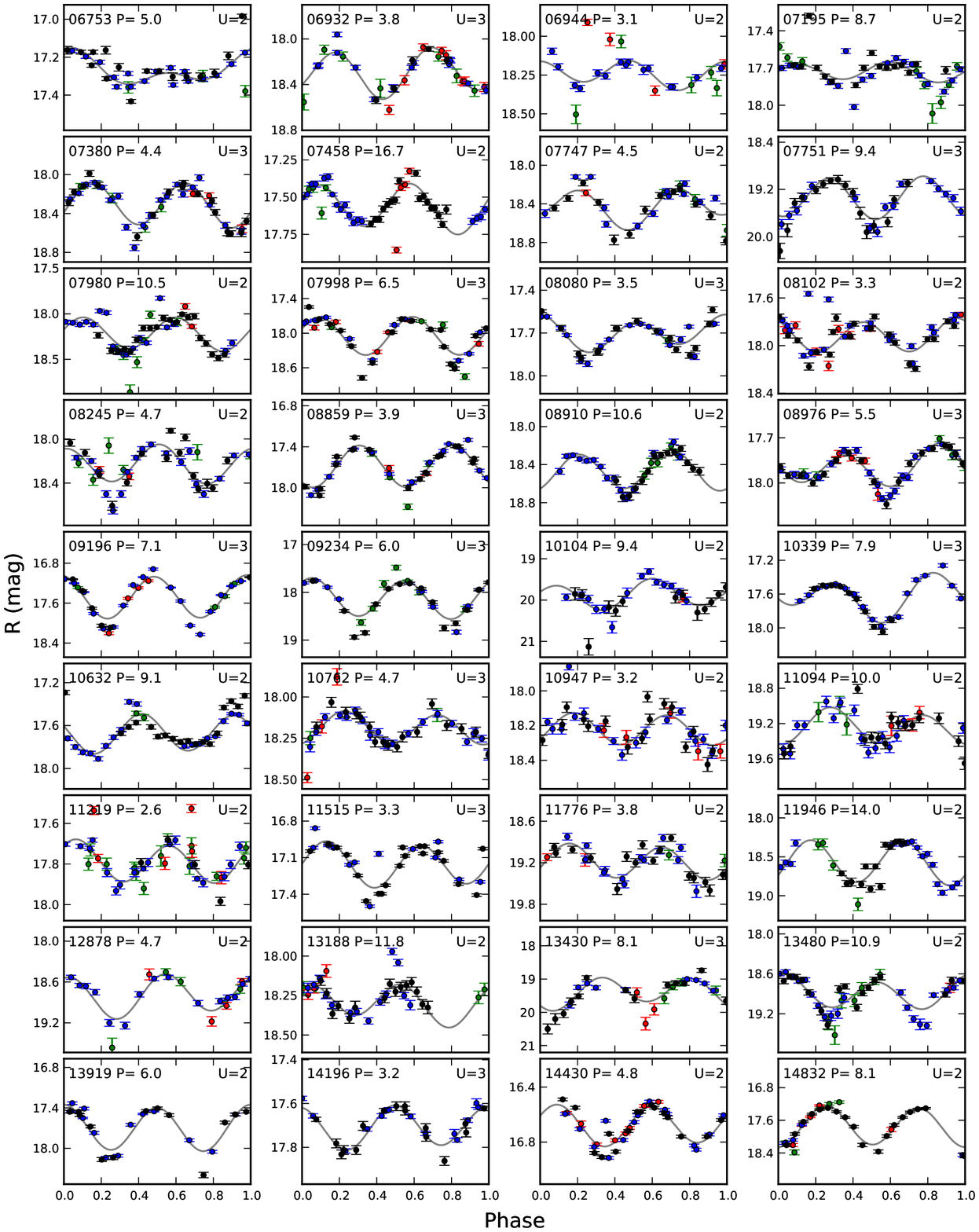}
  \caption{Same as Fig.~\ref{lightcurve00} for 36 more PTF-U2 asteroids.}
  \label{lightcurve01}
  \end{figure}

\begin{figure}
  %%\epsscale{0.8}
  \plotone{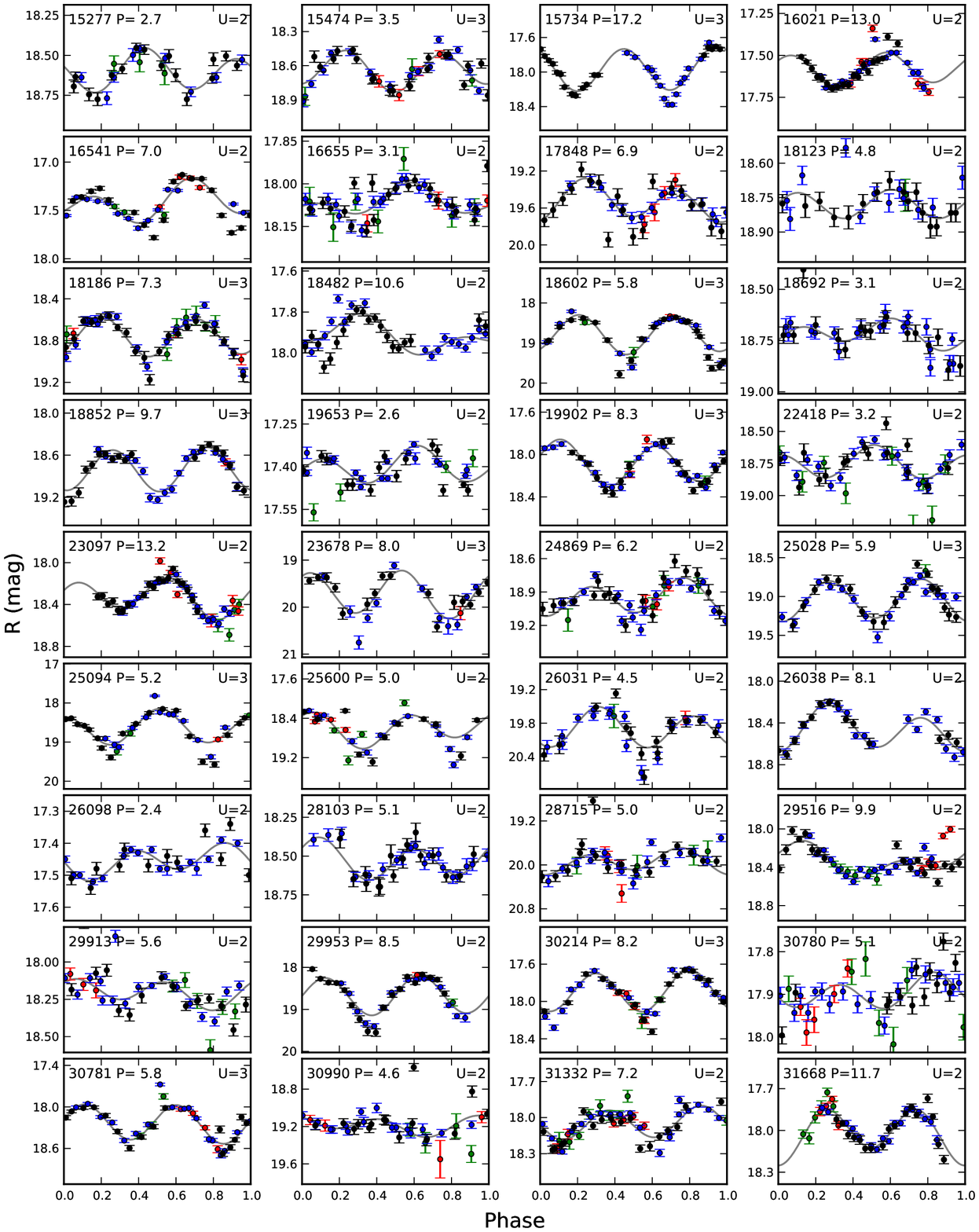}
  \caption{Same as Fig.~\ref{lightcurve00} for 36 more PTF-U2 asteroids.}
  \label{lightcurve02}
  \end{figure}

\begin{figure}
  %%\epsscale{0.8}
  \plotone{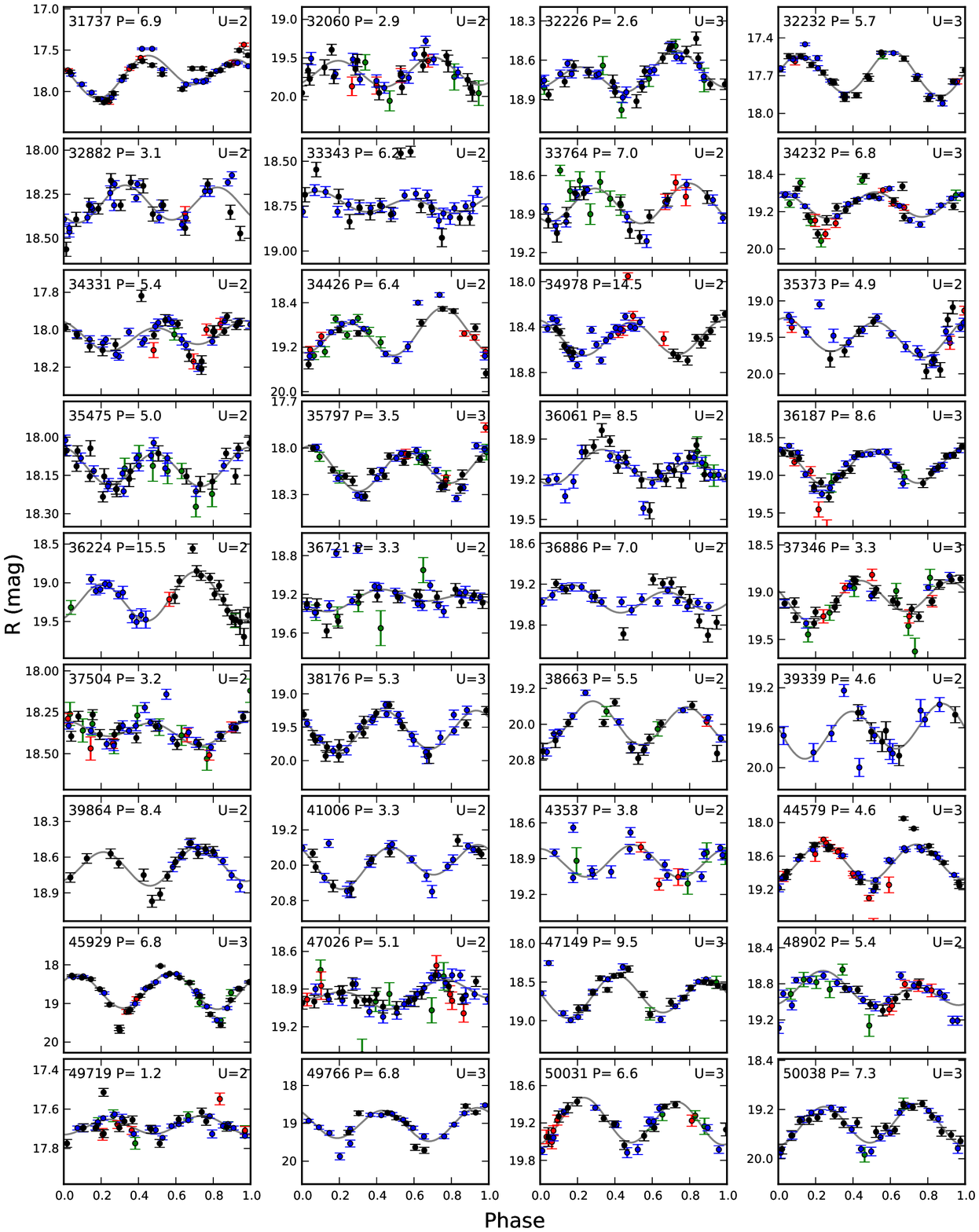}
  \caption{Same as Fig.~\ref{lightcurve00} for 36 more PTF-U2 asteroids.}
  \label{lightcurve03}
  \end{figure}

\begin{figure}
  %%\epsscale{0.8}
  \plotone{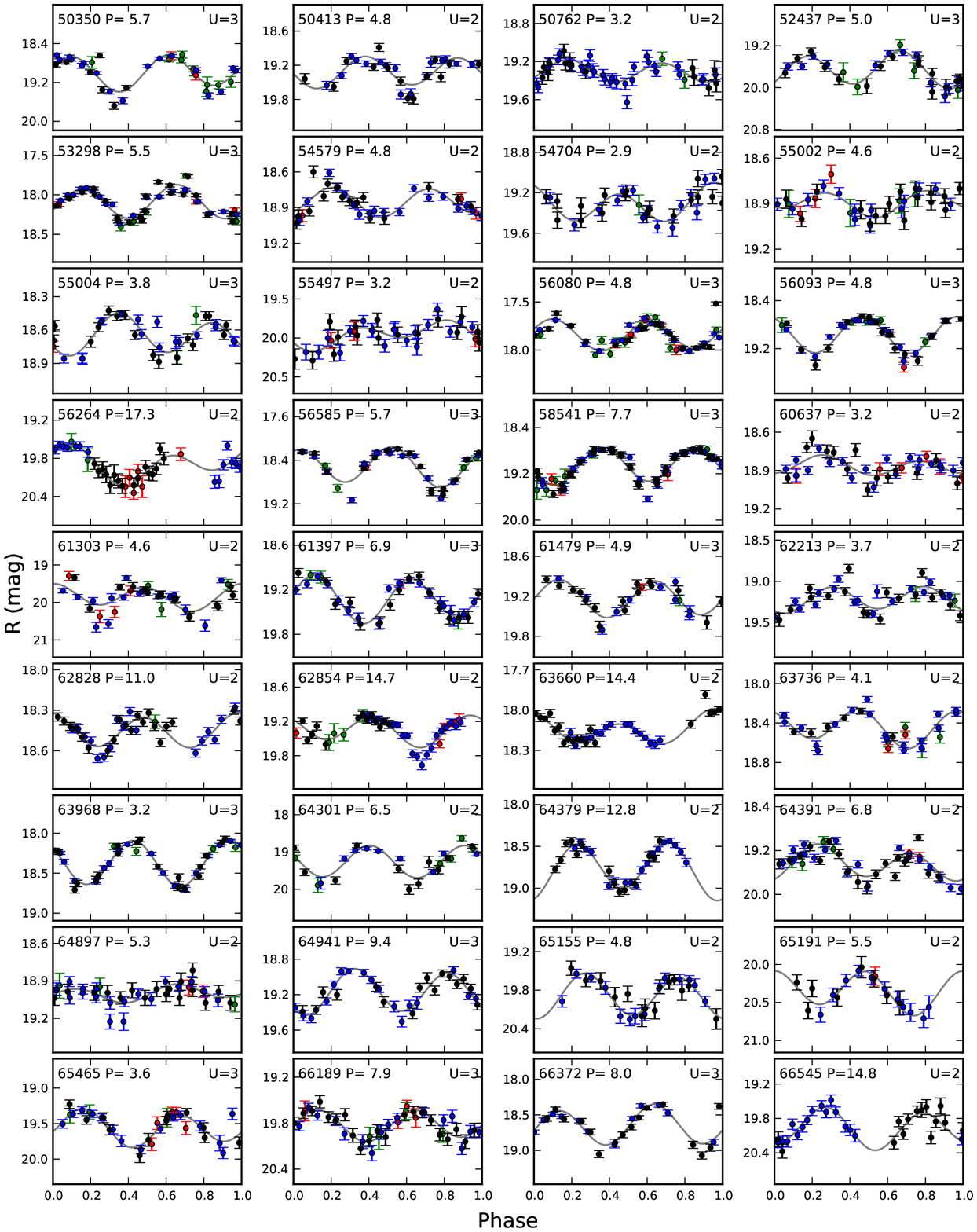}
  \caption{Same as Fig.~\ref{lightcurve00} for 36 more PTF-U2 asteroids.}
  \label{lightcurve04}
  \end{figure}

\begin{figure}
  %%\epsscale{0.8}
  \plotone{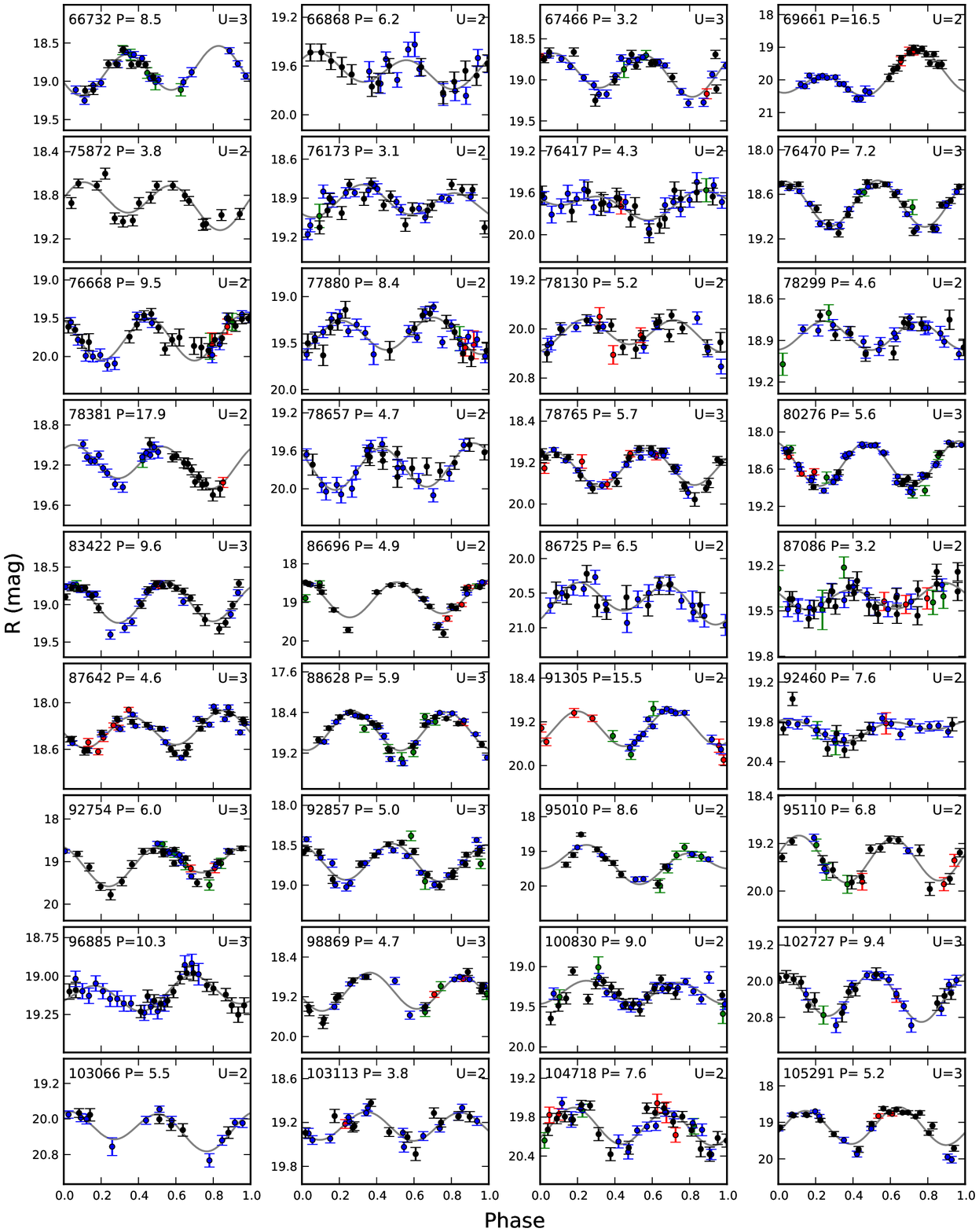}
  \caption{Same as Fig.~\ref{lightcurve00} for 36 more PTF-U2 asteroids.}
  \label{lightcurve05}
  \end{figure}

\begin{figure}
  %%\epsscale{0.8}
  \plotone{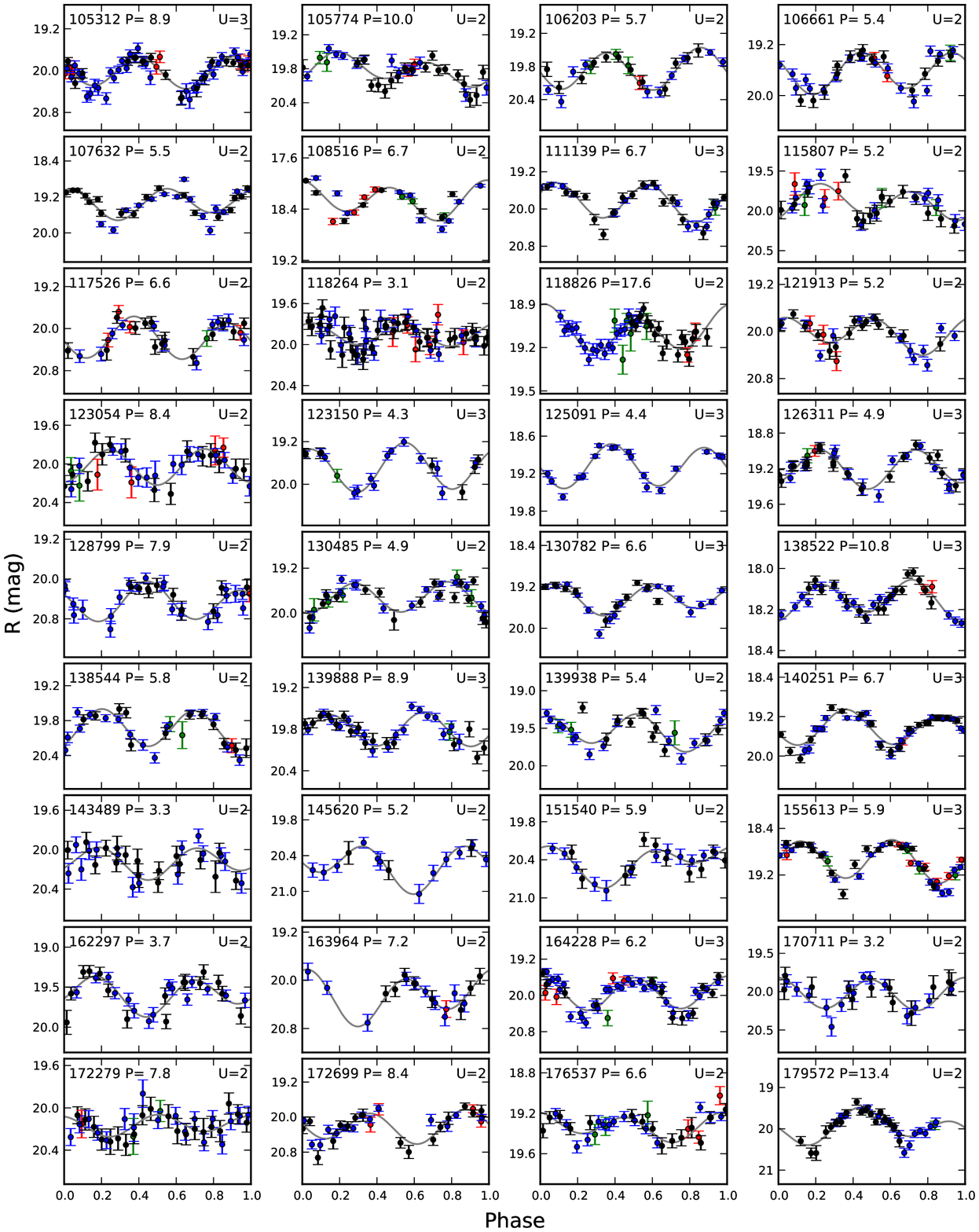}
  \caption{Same as Fig.~\ref{lightcurve00} for 36 more PTF-U2 asteroids.}
  \label{lightcurve06}
  \end{figure}

\begin{figure}
  %%\epsscale{0.8}
  \plotone{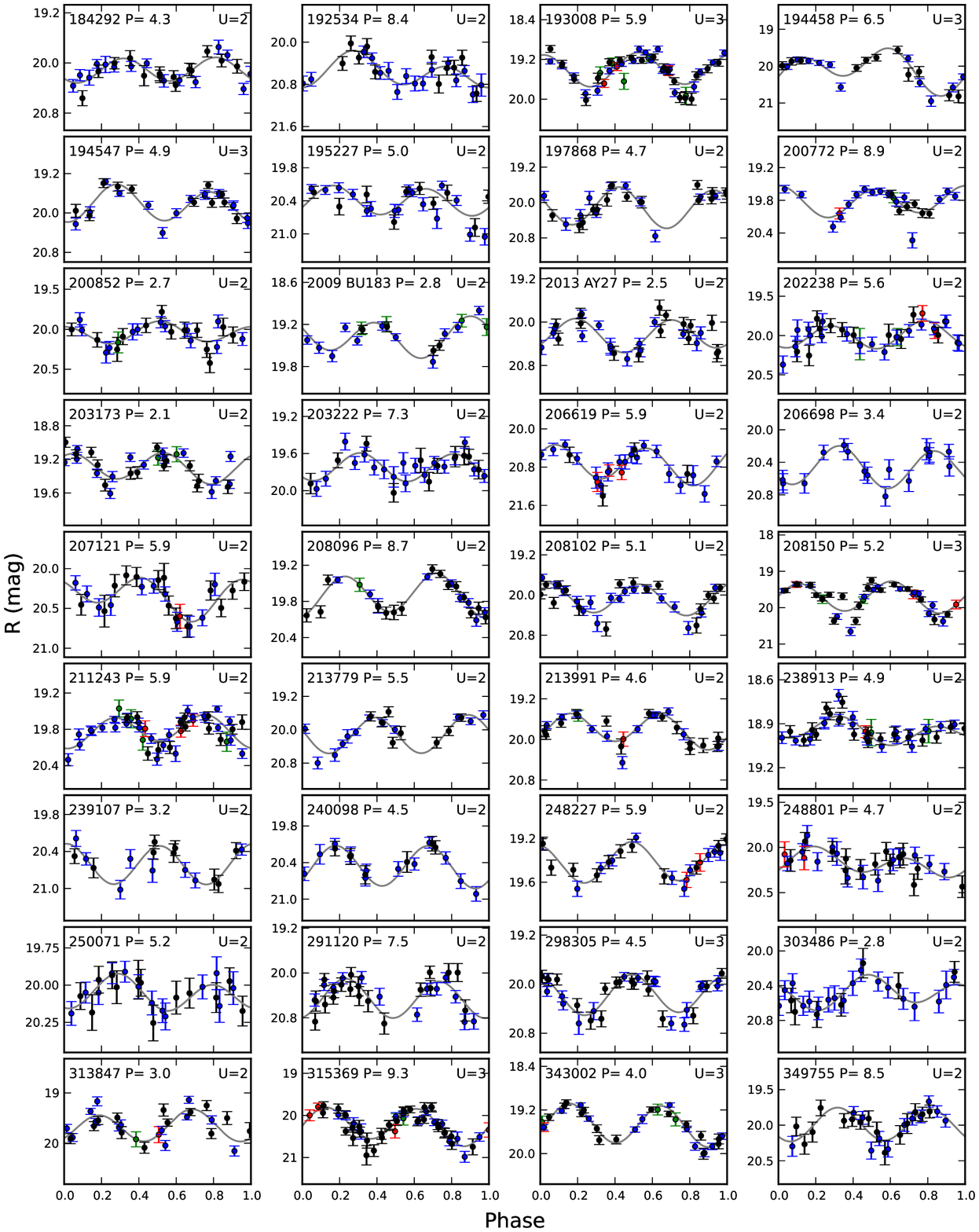}
  \caption{Same as Fig.~\ref{lightcurve00} for 36 more PTF-U2 asteroids.}
  \label{lightcurve07}
  \end{figure}

\begin{figure}
  %%\epsscale{0.8}
  \plotone{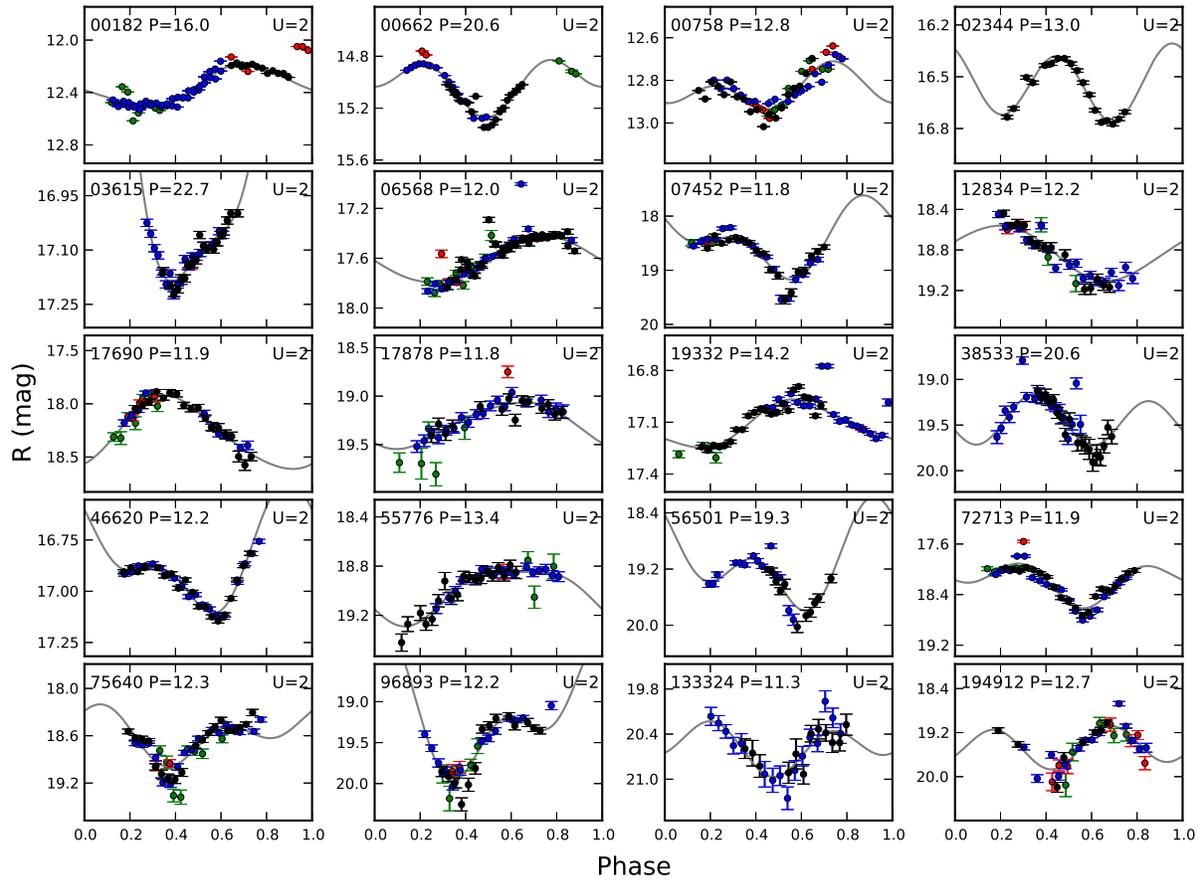}
  \caption{Same as Fig.~\ref{lightcurve00} for 20 more PTF-U2 asteroids,
  whose folded lightcurves only cover part of a rotation period.
  (7452) Izabelyuria and (75640) 2000 AE55 show binary asteroid features.}
  \label{lightcurve08}
  \end{figure}

\begin{figure}
  %%\epsscale{0.8}
  \plotone{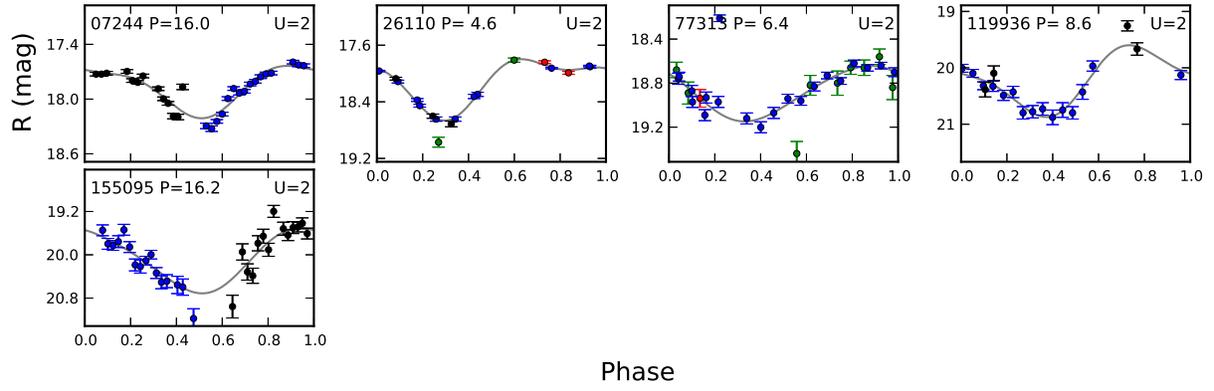}
  \caption{Same as Fig.~\ref{lightcurve00} for five more PTF-U2 asteroids,
  whose folded lightcurves show a single minimum.}
  \label{lightcurve09}
  \end{figure}

Among the 2500 PTF detected asteroids, 313 objects had rotation periods with $U \ge 2$ (hereafter, the
PTF-U2 asteroids). The information of the PTF-U2 asteroids are summarized in Table~\ref{table_p} and
their folded lightcurves are presented in Figs.~\ref{lightcurve00}--\ref{lightcurve09}. 18 PTF-U2
asteroids have published rotation periods in the Asteroid Light Curve Database
\citep[LCDB;][]{Warner2009}\footnote{http://www.minorplanet.info/lightcurvedatabase.html}, and 16 of
them show good agreements on their rotation periods that indicated that our result were highly
reliable. The possible reasons that the rotation periods of the other two asteroids, (182) Elsa and
(16541) 1991 PW18, showed discrepancies between this work and the LCDB are discussed below.

Compared to the published rotation period of $> 9$ hours with $U = 1$ for 1991 PW18
\citep{Warner2009}, ours is 7.0 hours with $U = 2$ which showed a clear folded lightcurve, therefore,
we believe our result to be more accurate.

The rotation period of (182) Elsa, a relatively large asteroid ($\sim40$ km), had been carefully
studied by \citet{Pilcher2009} with an intensively observed lightcurve. They gave a rotation period of
80.088 hours and a clear folded lightcurve that showed similar primary and secondary minimums. Since
our data set only covered part of its rotation period, the fitting procedure misidentifies a rotation
period of 15.97 hours (i.e., $\sim1/5$ of the published result) and gives a clear folded lightcurve.
As per the criteria described in Section \ref{period_analysis}, we still assigned $U = 2$ to our
rotation period.

From the case of (182) Elsa, we notice that when a folded lightcurve only covered part of a rotation
period, our fitting procedure was possible to lead us to an inaccurate rotation period. In 312 PTF-U2
asteroids, we found 20 objects faced this situation and five more showed folded lightcurves with a
single minimum. Since these 25 objects possibly had relatively large uncertainties in their rotation
periods, we exclude them in the following statistical analysis. The folded lightcurves of these 25
objects are presented in Fig.~\ref{lightcurve08} and ~\ref{lightcurve09}, respectively.

There were 49 PTF detected asteroids with $R <$ 18 mag and more than 20 detections which do not have
rotation period determinations (see Table~\ref{table_p_u0}). Since the small offset corrections for
the most PTF-U2 asteroids (i.e., $>$ 90\%) are of $R <$ 0.1 mag and the relative photometry accuracy
of $R = 18$ mag is $\sim$0.05 mag, it was unlikely to detect a lightcurve variation of $R <$ 0.1 mag
in our analysis and this has also been seen in Fig.~\ref{spin_amp}. Among 49 objects, only 5 belonged
to this case. The other 44 objects with lightcurve variations of $R >$ 0.1 mag should have had a
rotation period determination. However, most of these objects showed segmented lightcurves with a long
trend over our four-night observation time span (i.e., the lightcurve coverage was very limited). This
indicates that these objects might have relatively long rotation periods (i.e., several days) and
would have less chance to have rotation period determinations from our four-night observation. In
addition, 18 out of these 44 objects had diameters of $>$ 10 km and 6 of them (i.e., $\sim$33\%) have
lightcurve variations of $R >$ 0.3 mag. Comparing to the rest 26 objects, in which 10 objects
($\sim$38\%) have lightcurve variations of $R >$ 0.3 mag, both groups showed similar fractions for
lightcurve variation. However, this result was based on the lower limit of the lightcurve amplitude
(i.e., we only had partial lightcurve coverage) and should be verified when their full lightcurves are
available.

\subsection{Statistical Analysis}
\begin{figure}
  %%\epsscale{0.8}
  \plotone{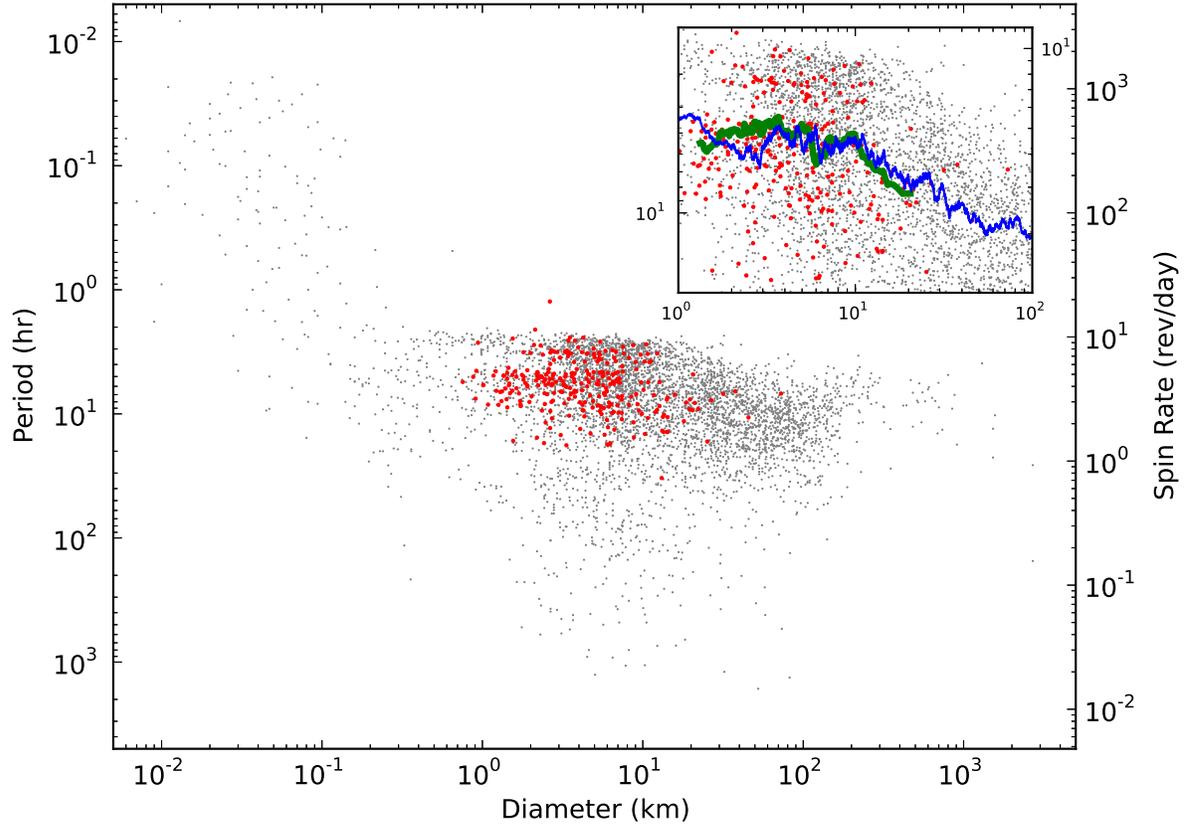}
  \caption{The plot of the diameters vs. rotation period. The red and gray filled circles are the PTF-U2
  asteroids and the LCDB objects with $U \geq 2$, respectively. The distribution of the PTF-U2 asteroids
  and that of the LCDB are similar. The ``spin barrier'' at $\sim2.2$ hours can obviously be seen for
  asteroids with diameters larger than a few hundred meters. The red filled circle above the ``spin barrier''
  is the super faster rotator candidate, (49719) 1999 VE50. The small plot at the upper-right corner is the
  detailed view of the dense region, where the green and blue lines are the geometric mean spin rates of
  the PTF-U2 asteroids and the LCDB, respectively}
  \label{dia_per}
  \end{figure}

\begin{figure}
  %%\epsscale{0.8}
  \plotone{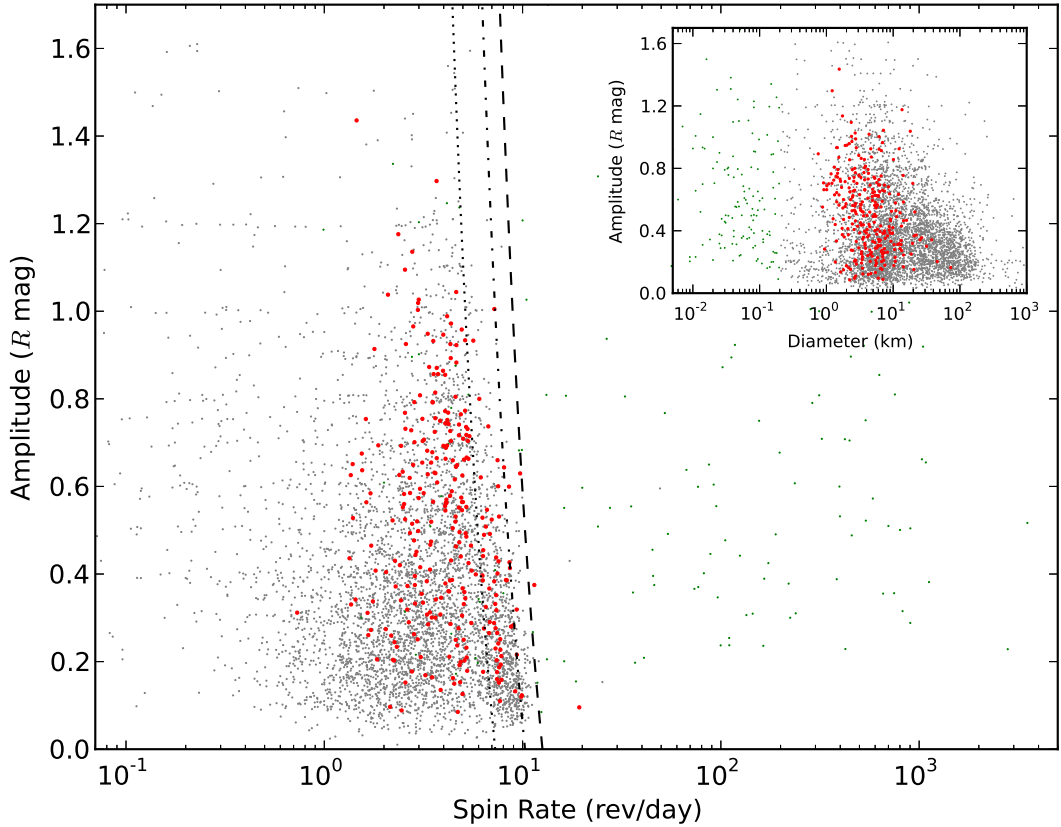}
  \caption{The plot of the spin rate vs. diameter. The red, gray and green filled circles are the PTF-U2
  asteroids, the LCDB objects with $D \geq 0.2$ km and the LCDB objects with $D < 0.2$ km, respectively.
  The dashed, dot-dashed and dotted lines represent the spin rate limits for ``rubble pile'' asteroids
  with bulk densities of 3, 2 and 1 g/cm$^3$ adopted from \citet{Pravec2000}.
  The small plot at the upper-right corner is the plot of the diameter vs. lightcurve amplitude for the PTF-U2 asteroids.}
  \label{spin_amp}
  \end{figure}

The asteroid spin rate limit is one of the particular interesting subjects. Fig.~\ref{dia_per} shows
the plot of the diameter vs. rotation period for the PTF-U2 asteroids and the objects adopted from the
Asteroid Light Curve Database with $U \geq 2$. The PTF-U2 asteroids occupy the dense region of the
plot. Because our observation was insensitive to detect a long rotation period, the PTF-U2 asteroids
showed a lack of slow rotators. The ``spin barrier'' at $\sim2.2$ hours can clearly be seen for
asteroids with diameters larger than a few hundred meters, which indicated the spin rate limit for
gravitationally bound aggregations (i.e., ``rubble pile''). Several super-fast-rotators (i.e.,
rotation period $<2.2$ hours) located at the upper-left corner, which were usually small sized objects
and could be of a monolithic nature \citep{Pravec2000}. The ``spin barrier'' can also been seen on
Fig.~\ref{spin_amp}, which shows the plot of the spin rate vs. lightcurve amplitude along with the
spin rate limits for ``rubble pile'' asteroids with bulk densities of 3, 2 and 1 g/cm$^3$ adopted from
\citet{Pravec2000}. One unusual object, (167714) 2001 OE84, which has a diameter of $\sim600$ m and a
rotation period of $\sim0.3$ hours faster than the ``spin barrier'', and \citet{Pravec2002} treated
this object as an exceptional case. However, \citet{Holsapple2007} introduced a size-dependent
strength asteroid model, which included tensile and cohesiveness in addition to gravity, to explain
the existence of kilometer size super-fast-rotators. In this study, we detected one super-fast-rotator
candidate, (49719) 1999 VE50, which has a rotation period of 1.24 hours, an amplitude of $\sim0.29$
mag and a diameter of $\sim2.6$ km. However, as pointed out by \citet{Harris2012}, this could be a
result of a random combination of scattered data points showing relatively small lightcurve amplitude.
Therefore, to confirm such fast spin rate of 1999 VE50 requires a followup observation and a detailed
investigations.

The small plot at the upper right corner of Fig.~\ref{dia_per} is the detailed view of the dense
region, on which we also plot the geometric mean spin rate of the PTF-U2 asteroids and that of the
LCDB by using a running box containing 30 and 100 objects, respectively. Both geometric mean spin
rates begin with a flat and then start to decrease at $D \sim 10$ km. The decrease at $D \sim 10$ km
was considered as a consequence of the transition from the ``small'' asteroids, which showed a
non-Maxwellian spin rate distribution with an excess of both fast and slow rotators, to the ``large''
asteroids, which showed a Maxwellian spin rate distribution \citep{Pravec2002}. Several studies
indicated the ``large'' asteroid begins at the 30--50 km diameter range \citep{Fulchignoni1995,
Donnison1999, Pravec2000}, but we were not able to determine where the ``large'' asteroids begins due
to only a few objects with $D > 30$ km in the PTF-U2 asteroids.

\begin{figure}
  %%\epsscale{0.8}
  \plotone{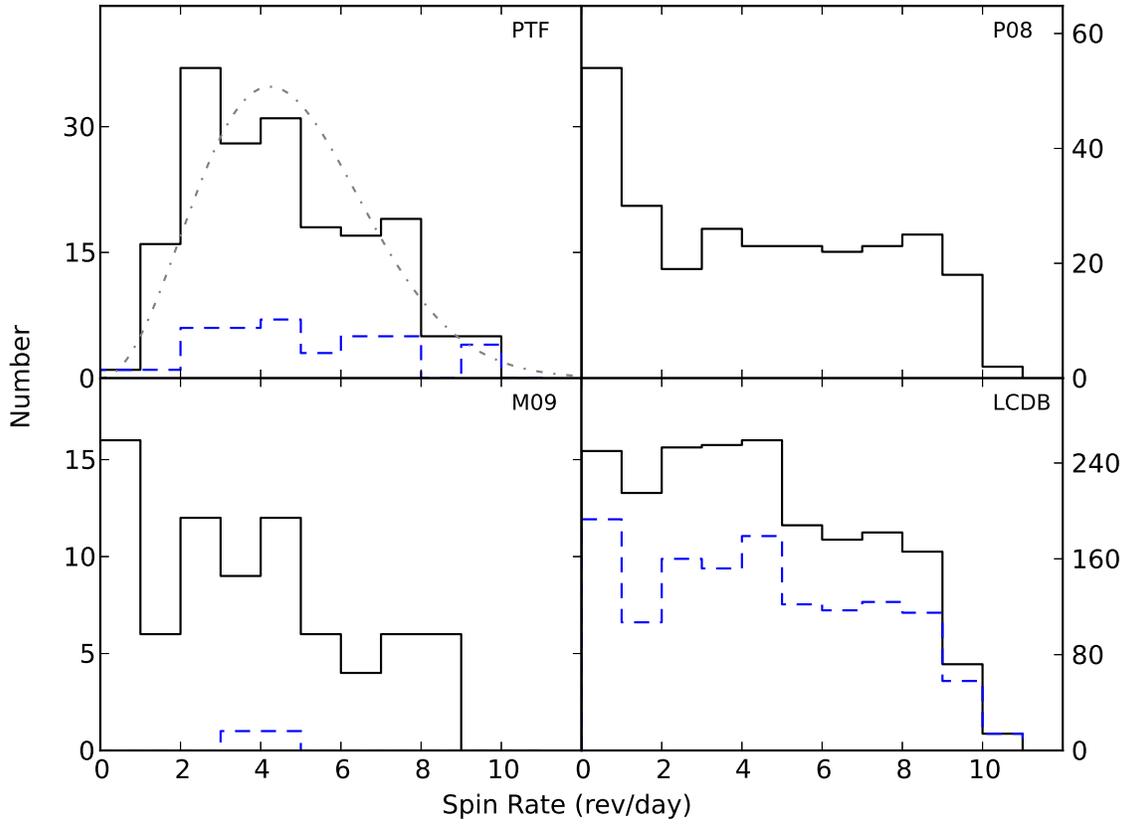}
  \caption{The distributions of spin rate of asteroids with $3 < D \leq 15$ km (solid lines) for the PTF-U2 asteroids
  (upper-left), P08 (upper-right), M09 (lower-left) and the LCDB. The dashed lines are the asteroids with
  $a < 2.5$ AU. The dot-dashed line on the PTF-U2 asteroids is the best-fit Maxwellian distribution.}
  \label{spin_rate_comp}
  \end{figure}

The spin rate distribution of the ``small'' asteroids also provided valuable information to understand
asteroid spin rate evolution. In order to compare with the result of \citet[][P08
hereafter]{Pravec2008}, we selected asteroids with 3 km $< D \le$ 15 km from the PTF-U2 asteroids,
\citep[][M09 hereafter]{Masiero2009} and the LCDB to generate spin rate distributions in
Fig.~\ref{spin_rate_comp}. In contrast to the flat distribution of P08, the other three showed a
number decrease at the spin rate of $>$ 5 rev/day. \citet{Warner2009} pointed out the number decrease
was an observational bias due to a tendency toward smaller lightcurve amplitude with increasing spin
rates (see Fig.~\ref{spin_amp}), while M09 showed the multiple YORP-braking stages that existed in
main belt asteroids could lead to such a result. However, M09 also mentioned that the discrepancy
between P08 and the others could be the result of different survey methods (i.e., the PTF, M09 and the
LCDB are untargeted surveys or collecting data set while P08 targets individual asteroids with $a <
2.5$ AU). If we restrict the comparison to the asteroids with $a < 2.5$ AU, regardless of only two
samples in M09, the number decrease at the spin rate of $>$ 5 rev/day still showed on the PTF-U2
asteroids and the LCDB (see the dashed lines in Fig.~\ref{spin_rate_comp}). Therefore, a large sample
of asteroid rotation period from a single survey like the PTF is of paramount importance to reveal the
spin rate distribution of the ``small'' asteroids. The best-fit Maxwellian distribution for the PTF-U2
asteroids is also shown on Fig.~\ref{spin_rate_comp}. We see the PTF-U2 asteroids deviate somewhat to
the Maxwellian form, which means that mechanisms, other than collision, involved in asteroid spin rate
evolution as well.

\begin{figure}
  %%\epsscale{0.5}
  \plottwo{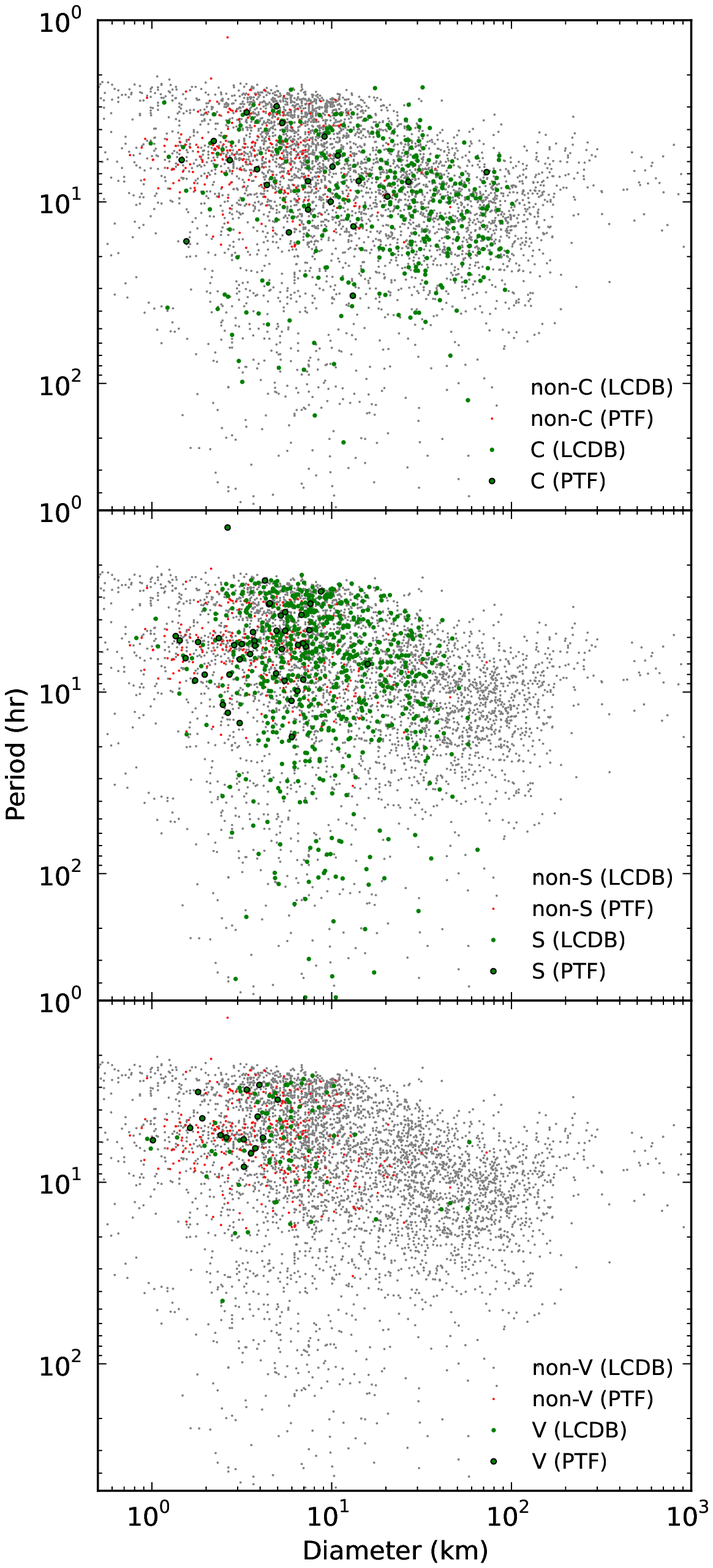}{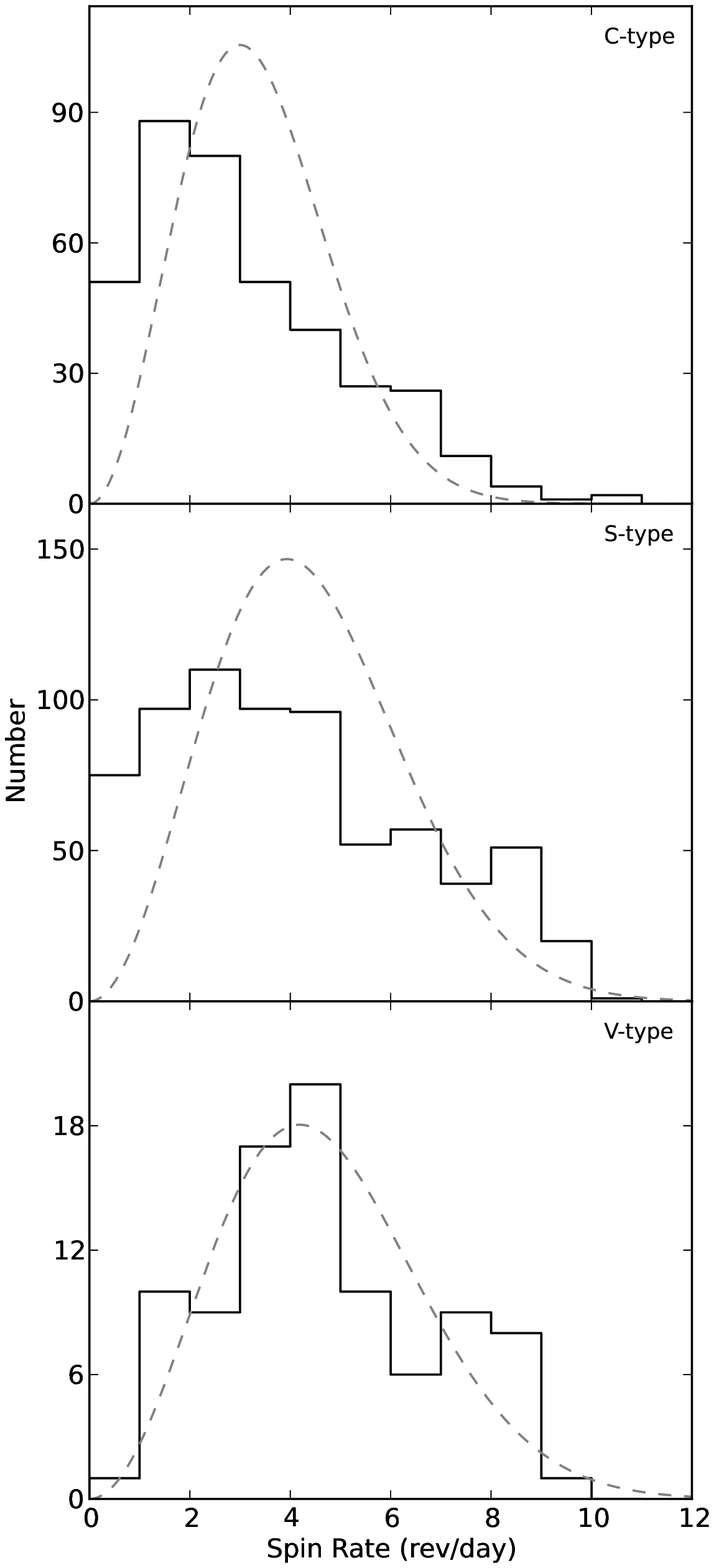}
  \caption{Left column: The plots of the diameter vs. rotation period for C- (upper panel), S- (middle panel) and
  V-type (lower panel) asteroids. The bigger green filled circles with black edge and the smaller green
  filled circles represent PTF-U2 asteroids and the LCDB objects with available taxonomy, respectively.
  The red and gray dots are the PTF-U2 asteroids and LCDB objects other than the corresponding taxonomic
  type in each plot, respectively. The taxonomic types are obtained from SDSS colors.
  Right column: The distributions of spin rate (solid lines) for C- (upper panel), S- (middle panel)
  and V-type (lower panel) asteroids. The dashed line on each panel is the best-fit Maxwellian distribution.
  The samples include both the PTF-U2 asteroids and the LCDB objects.}
  \label{dia_per_tax}
  \end{figure}

To gain an approximate idea of how the spin rate distributes for different taxonomic type asteroids,
we summarized the plots of the diameter vs. rotation period and the distributions of spin rate for
available S-, C- and V-type asteroids in Fig.~\ref{dia_per_tax}. The taxonomic types were determined
by using SDSS color \citep{Parker2008}. We were not able to tell the difference between the scatter
plots from one another by eyes (the left column in Fig.~\ref{dia_per_tax}), however, the spin rate
distributions showed some difference (the right column in Fig.~\ref{dia_per_tax}). The C-type's
distribution showed a decreasing number with increasing spin rate, the S-type's distribution had a
number drop at spin rate of $> 5$ rev/day and the V-type's distribution demonstrated a number
enhancement around spin rate of 6--5 rev/day. Moreover, none of them could be fitted reasonably by a
Maxwellian distribution. When we ran the two-sample Kolmogorov-Smirnov (KS) test for each pair of
these three type asteroids to compare their spin rate distributions, all the KS test $p-$values were
much lower than 0.005. Although this indicates these three different types were unlikely to come from
the same population, we believed it could be greatly affected by insufficient number and
incompleteness in our samples, especially the different diameter ranges of these three subgroups (see
the left column in Fig.~\ref{dia_per_tax}). Therefore, we expect to have a more accurate comparison on
this topic when more asteroid rotation periods are available from observations now in planning.

The asteroid lightcurve profile is a powerful tool to probe asteroid shape as well as to discover
binary asteroid. The inset of Fig.~\ref{spin_amp} shows the plot of the diameter vs. lightcurve
amplitude. Both the PTF-U2 asteroids and the LDCB show a boundary from the upper-middle to the
lower-right that indicates relatively large asteroids tend to have small lightcurve amplitudes. Most
PTF-U2 asteroids show simple sinusoidal-like folded lightcurves. However, we found two binary asteroid
candidates, (7452) Izabelyuria and (75640) 2000 AE55 (see Fig.~\ref{lightcurve08}), whose lightcurves
showed a deep V-shape minima and an inverse U-shape maxima with a relatively long rotation period
\citep{Pravec2006}. The fractions of binary asteroid in different asteroid groups provide important
constraints on the binary asteroid formation models. Therefore, an overall survey on asteroid binary
population with large samples could reveal a clearer picture on how different mechanisms work on
binary asteroid formation.

\section{Summary and Discussion}
This study has demonstrated the capability of the PTF to pursue large amount of asteroid rotation
periods via a 12-PTF-fields survey in a four consecutive nights observation with a cadence of $\sim20$
min during February 15--18, 2013. The PTF photometric data were used to extract the lightcurves of
known asteroids and measure their rotation periods. There were 2500 known asteroids with more than
five detections and 312 of them had highly reliable rotation periods in this work. The plots of the
spin rate vs. diameter for the PTF-U2 asteroids and the LCDB were very similar. Both show the ``spin
barrier'' clearly at $\sim2.2$ hours and similar geometric mean spin rates. The spin rate distribution
of the PTF-U2 asteroids showed a number decrease at the spin rate of $>$ 5 rev/day, which is not a
Maxwellian distribution nor as flat as shown by \citet{Pravec2008}. The non-Maxwellian distribution
indicated that the asteroid spin rate evolution was not only affected by collision, but also by other
mechanics, in particular the YORP effect. The rough test for available S-, C- and V-type asteroids
does not have a clear conclusion on any difference between the spin rate distributions of various
taxonomic types. However, we did find one super-fast-rotator candidate, (49719) 1999 VE50, which had a
rotation period of 1.24 hours and a diameter of $\sim2.6$ km. If the fast spin rate of 1999 VE50 is
confirmed, the size-dependent strength asteroid model will be supported. In addition, we also detected
two binary asteroid candidates, (7452) Izabelyuria and (75640) 2000 AE55. The tendency toward smaller
lightcurve amplitudes with increasing diameter is seen in the PTF-U2 asteroids as well, which means
large asteroids tend to have rounder shapes.

Since our target fields are close to the Galactic plane toward the anti-Galactic center direction, the
detection rate of known asteroids is expected to be lower than that in the off-Galactic plane fields,
and so does the amount of highly reliable rotation period. For an approximate estimation, it is very
likely to obtain 10000 asteroid rotation periods by reproducing this kind of observation around 20--30
times. In addition, we also expect to retrieve more asteroid rotation periods from previously observed
high cadence fields, which is now under analysis. Such a huge amount of asteroid rotation periods can
provide more definite constraints on various studies, such as how different mechanisms (e.g.,
collision, YORP effect and tidal force during their encounter with planets) involvee in asteroid spin
rate evolution, how tensile and cohesiveness account for the spin rate limits of different size
asteroids \citep[i.e., the size-dependent strength for asteroids;][]{Holsapple2007} and the censor on
asteroid lightcurve profile (i.e., asteroid shape/axis ratio). We also can further investigate the
asteroid spin rate as a function of the size, taxonomy, dynamical group and asteroid family. In
addition, more binary asteroids will be discovered to reveal their fractions in different asteroid
populations and provide important constraints on their formation models.

\acknowledgments This work is supported in part by the National Science Council of Taiwan under the
grants NSC 101-2119-M-008-007-MY3 and NSC 102-2112-M-008-019-MY3. We would like to thank the valueable
comments and suggestions from Eran Ofek, which makes the manuscript more complete. We also thank the
useful comments from the referee, Joseph Masiero, which have helped to improve the content of the
paper. This publication makes use of data products from WISE, which is a joint project of the
University of California, Los Angeles, and the Jet Propulsion Laboratory/California Institute of
Technology, funded by the National Aeronautics and Space Administration. This publication also makes
use of data products from NEOWISE, which is a project of the Jet Propulsion Laboratory/California
Institute of Technology, funded by the Planetary Science Division of the National Aeronautics and
Space Administration. We gratefully acknowledge the extraordinary services specific to NEOWISE
contributed by the International Astronomical Union¡¦s Minor Planet Center, operated by the
Harvard-Smithsonian Center for Astrophysics, and the Central Bureau for Astronomical Telegrams,
operated by Harvard University.

\begin{deluxetable}{lrrccccc}
\tabletypesize{\scriptsize} \tablecaption{The observation information. \label{obs_log}}
\tablewidth{0pt} \tablehead{ \colhead{Field ID} & \colhead{RA} & \colhead{Dec.} & \colhead{[$l$, $b$]}
& \colhead{Feb 15} &
\colhead{Feb 16} & \colhead{Feb 17} & \colhead{Feb 18} \\
\colhead{} & \colhead{($^{\circ}$)} & \colhead{($^{\circ}$)} & \colhead{($^{\circ}$)} &
\colhead{$\Delta$t, N$_\textrm{exp}$} & \colhead{$\Delta$t, N$_\textrm{exp}$} & \colhead{$\Delta$t,
N$_\textrm{exp}$} & \colhead{$\Delta$t, N$_\textrm{exp}$}} \startdata
    3654 &  101.02 &   21.38 &  [192.97, ~8.02] &  3.4, 10 &   5.1, 16 &   6.8, 19 &   6.4, 19 \\
    3655 &  104.69 &   21.38 &  [194.47, 11.11] &  3.4, 10 &   5.3, 15 &   6.8, 19 &   6.4, 19 \\
    3656 &  108.37 &   21.38 &  [195.92, 14.23] &  3.4, 10 &   5.3, 15 &   7.2, 20 &   6.7, 20 \\
    3657 &  112.04 &   21.38 &  [197.33, 17.36] &  3.4, 10 &   5.4, 15 &   7.2, 20 &   7.1, 21 \\
    3658 &  115.71 &   21.38 &  [198.72, 20.52] &  3.4, 10 &   5.5, 16 &   7.6, 21 &   7.4, 22 \\
    3749 &   90.93 &   23.62 &  [186.63, ~0.81] &  3.4, 10 &   5.3, 14 &   6.4, 18 &   6.0, 18 \\
    3750 &   94.64 &   23.62 &  [188.27, ~3.79] &  3.4, 10 &   5.1, 15 &   6.4, 18 &   6.0, 18 \\
    3751 &   98.35 &   23.62 &  [189.84, ~6.81] &  3.4, 10 &   5.1, 15 &   6.8, 19 &   6.4, 19 \\
    3752 &  102.06 &   23.62 &  [191.34, ~9.87] &  3.4, 10 &   5.0, 14 &   6.8, 19 &   6.7, 20 \\
    3753 &  105.77 &   23.62 &  [192.80, 12.95] &  3.4, 10 &   5.1, 15 &   7.2, 20 &   6.7, 20 \\
    3754 &  109.48 &   23.62 &  [194.21, 16.07] &  3.4, 10 &   5.1, 15 &   7.6, 21 &   7.1, 21 \\
    3755 &  113.20 &   23.62 &  [195.58, 19.21] &  3.4, 10 &   5.4, 15 &   7.6, 21 &   7.4, 22 \\
\enddata
\tablecomments{$\Delta$t is the time duration spanned by each observing set in hours and
N$_\textrm{exp}$ is the total number of exposures for each night and field.}
\end{deluxetable}

\begin{deluxetable}{lllllllllllllllllll}
\tabletypesize{\scriptsize} \rotate \tablecaption{Synodic rotation periods of 313 asteroids with $U
\geq 2$. \label{table_p}} \tablewidth{0pt} \tablehead{ \colhead{Obj ID} & \colhead{Designation} &
\colhead{$a$} & \colhead{$e$} & \colhead{$i$} & \colhead{$\Omega$} & \colhead{$\omega$} &
\colhead{Epoch} & \colhead{D} & \colhead{$\triangle$} & \colhead{$r$} & \colhead{$\alpha$} &
\colhead{$H$} & \colhead{n} & \colhead{m} & \colhead{$PTF_R$} & \colhead{Period (hr)} &
\colhead{$\triangle m$} & \colhead{U}} \startdata
00182\tablenotemark{*\dag}     & (182) Elsa                   &  2.42 &  0.19 &  2.00 & 107.2 & 310.2 & K13B4 &  44.0\tablenotemark{w}   &   2.18 &   1.32 & 16.50 &  9.07$\pm$0.33   &  4 &  84 & 12.38$\pm$0.00   &  15.97$\pm$0.31  &  0.34 &  2 \\
00300\tablenotemark{*}     & (300) Geraldina              &  3.21 &  0.06 &  0.73 &  42.7 & 327.1 & K134I &  72.7\tablenotemark{w}   &   3.23 &   2.52 & 13.82 &  9.51$\pm$0.09   &  4 &  57 & 14.85$\pm$0.00   &   6.86$\pm$0.02  &  0.16 &  3 \\
00543\tablenotemark{*}     & (543) Charlotte              &  3.06 &  0.16 &  8.48 & 295.0 & 109.8 & K134I &  45.6\tablenotemark{w}   &   2.87 &   2.06 & 13.24 &  9.23$\pm$0.09   &  4 &  61 & 13.85$\pm$0.00   &  10.69$\pm$0.05  &  0.20 &  3 \\
00662\tablenotemark{*\dag}     & (662) Newtonia               &  2.55 &  0.22 &  4.12 & 133.7 & 165.7 & K134I &  22.4\tablenotemark{w}   &   3.11 &   2.29 & 12.05 & 10.00$\pm$0.15   &  4 &  47 & 15.06$\pm$0.00   &  20.60$\pm$0.18  &  0.42 &  2 \\
00758\tablenotemark{*\dag}     & (758) Mancunia               &  3.19 &  0.15 &  5.61 & 106.2 & 315.2 & K134I &  85.0\tablenotemark{w}   &   2.87 &   2.09 & 14.07 &  8.11$\pm$0.16   &  4 &  62 & 12.85$\pm$0.00   &  12.77$\pm$0.05  &  0.26 &  2 \\
01350\tablenotemark{*}     & (1350) Rosselia              &  2.86 &  0.09 &  2.94 & 139.6 & 238.5 & K134I &  21.1\tablenotemark{w}   &   2.86 &   2.12 & 15.20 & 10.36$\pm$0.25   &  4 &  59 & 14.99$\pm$0.00   &   8.16$\pm$0.01  &  0.52 &  3 \\
01633\tablenotemark{*}     & (1633) Chimay                &  3.19 &  0.12 &  2.68 & 114.1 &  65.0 & K134I &  37.7\tablenotemark{w}   &   2.96 &   2.16 & 13.19 & 10.36$\pm$0.17   &  4 &  55 & 15.21$\pm$0.00   &   6.58$\pm$0.01  &  0.34 &  3 \\
01778     & (1778) Alfven                &  3.15 &  0.13 &  2.47 & 106.2 & 135.8 & K134I &  20.6\tablenotemark{w}   &   3.33 &   2.51 & 11.15 & 11.59$\pm$0.15   &  4 &  55 & 16.88$\pm$0.01   &   4.82$\pm$0.00  &  0.37 &  3 \\
\enddata
\tablecomments{Columns: asteroid's designations, semimajor axis ($a$, AU), eccentricity ($e$, degree),
inclicnation ($i$, degree), longitude of the ascending node ($\Omega$, degree), argument of periapsis
($\omega$, degree), epoch, derived diameter (D, km), heliocentric distance ($\triangle$, AU), geodesic
distance ($r$, AU), phase angle ($\alpha$, degree), absolute magnitude ($H$, mag), number of nights
(n), number of images (m), $PTF_R$ magnitude, derived rotation period (hours), lightcurve amplitude
(mag) and rotation period quality code (U). The amplitudes of the objects of partial lightcurve
coverage and single minimun lightcurve should be treated as low limits.} \tablenotetext{*}{The
asteroids have published rotation periods.}\tablenotetext{\dag}{The objects of partial lightcurve
coverage.} \tablenotetext{\vee}{The objects of single minimum lightcurve.} \tablenotetext{w}{The
$WISE$/NEOWISE diameter.}
\end{deluxetable}

\begin{deluxetable}{lllllllllllllllllll}
\tabletypesize{\scriptsize} \rotate \tablecaption{The 49 bright PTF detected asteroids with $U \leq
1$. \label{table_p_u0}} \tablewidth{0pt} \tablehead{ \colhead{Obj ID} & \colhead{Designation} &
\colhead{$a$} & \colhead{$e$} & \colhead{$i$} & \colhead{$\Omega$} & \colhead{$\omega$} &
\colhead{Epoch} & \colhead{D} & \colhead{$\triangle$} & \colhead{$r$} & \colhead{$\alpha$} &
\colhead{$H$} & \colhead{n} & \colhead{m} & \colhead{$PTF_R$} & \colhead{Period (hr)} &
\colhead{$\triangle m$} & \colhead{U}} \startdata
01004     & (1004) Belopolskya           &  3.40 &  0.09 &  2.98 & 153.6 & 216.3 & K134I &  74.0   &   3.44 &   2.71 & 12.56 &  9.41$\pm$0.08   &  4 &  63 & 15.00$\pm$0.00   &   9.06$\pm$0.03  &  0.12 &  1 \\
01780     & (1780) Kippes                &  3.02 &  0.05 &  9.00 & 291.0 & 339.8 & K134I &  31.3\tablenotemark{w}   &   3.17 &   2.54 & 15.39 & 10.15$\pm$0.06   &  4 &  59 & 15.53$\pm$0.01   &                &  0.11 &    \\
01986     & (1986) Plaut                 &  3.09 &  0.20 &  2.21 & 146.6 & 230.8 & K134I &  19.6\tablenotemark{w}   &   3.02 &   2.28 & 14.37 & 11.82$\pm$0.07   &  3 &  31 & 16.83$\pm$0.01   &  13.62$\pm$0.54  &  0.16 &  1 \\
02264     & (2264) Sabrina               &  3.13 &  0.17 &  0.16 & 240.3 &  79.1 & K134I &  37.3\tablenotemark{w}   &   3.60 &   2.85 & 11.62 & 10.50$\pm$0.10   &  4 &  50 & 16.30$\pm$0.01   &                &  0.19 &    \\
02273     & (2273) Yarilo                &  2.45 &  0.17 &  0.39 &  66.1 &  41.5 & K134I &   5.6\tablenotemark{w}   &   2.06 &   1.21 & 18.71 & 13.50$\pm$0.07   &  4 &  65 & 16.46$\pm$0.01   &                &  0.10 &    \\
02668     & (2668) Tataria               &  2.32 &  0.08 &  3.16 & 298.2 &  62.9 & K134I &   5.4\tablenotemark{w}   &   2.37 &   1.69 & 20.82 & 13.22$\pm$0.25   &  4 &  50 & 17.24$\pm$0.02   &                &  0.53 &    \\
03187     & (3187) Dalian                &  2.28 &  0.06 &  2.76 & 312.4 & 105.1 & K134I &   6.2\tablenotemark{w}   &   2.21 &   1.44 & 19.78 & 13.02$\pm$0.11   &  4 &  54 & 16.53$\pm$0.01   &                &  0.22 &    \\
03462     & (3462) Zhouguangzhao         &  2.45 &  0.21 &  5.80 & 113.1 & 254.9 & K134I &   5.5   &   2.46 &   1.77 & 19.41 & 13.32$\pm$0.40   &  4 &  42 & 17.42$\pm$0.02   &                &  0.75 &    \\
\enddata
\tablecomments{Columns: asteroid's designations, semimajor axis ($a$, AU), eccentricity ($e$, degree),
inclicnation ($i$, degree), longitude of the ascending node ($\Omega$, degree), argument of periapsis
($\omega$, degree), epoch, derived diameter (D, km), heliocentric distance ($\triangle$, AU), geodesic
distance ($r$, AU), phase angle ($\alpha$, degree), absolute magnitude ($H$, mag), number of nights
(n), number of images (m), $PTF_R$ magnitude, derived rotation period (hours), lightcurve varition
(mag) and rotation period quality code (U).} \tablenotetext{w}{The $WISE$/NEOWISE diameter.}
\end{deluxetable}

\clearpage

%%  \begin{figure}
%%  %%\epsscale{0.8}
%%  \plotone{FT_aeih01_diffphoidx_pcen5.eps}
%%  \caption{Scatter plots for the PTF detected known objects. From upper-left to lower-right panels are eccentricity vs. semimajor axis,
%%  inclination vs. semimajor axis, absolute magnitude vs. semimajor axis, inclination vs. eccentricity, absolute magnitude vs. inclination,
%%  and absolute magnitude vs. eccentricity}
%%  \label{aeih_dist01}
%%  \end{figure}

\clearpage
%%  \begin{figure}
%%  %%\epsscale{0.8}
%%  \plotone{FT_period_amp_diffphoidx_pcen5.eps}
%%  \caption{The plots for derived diameters vs. rotation periods (upper-left) and derived diameters vs. lightcurve amplitudes (upper-right);
%%  and the number distributions for rotation periods (lower-left, 2 hours bins) and lightcurve amplitudes (lower-right, 0.05 mag bibs).
%%  All data points shown here are PTF detected known objects with rotation period quality code $U\geq2$.}
%%  \label{per_amp}
%%  \end{figure}

%%  \begin{figure}
%%  %%\epsscale{0.8}
%%  \plotone{FT_dia_amp_diffphoidx_pcen5.eps}
%%  \caption{The plot of the diameter vs. lightcurve amplitude for the PTF-U2 asteroids.}
%%  \label{amp}
%%  \end{figure}

\clearpage


\begin{thebibliography}{}
\bibitem[Ag{\"u}eros et al.(2011)]{Agueros2011} Ag{\"u}eros, M.~A., Covey, K.~R., Lemonias, J.~J., et al.\ 2011, \apj, 740, 110
%\bibitem[Behrend(2006)]{Behrend2006} Behrend, R.\ 2006, Observatoire de Geneve website, http://obswww.unige.ch/~behrend/page\_cou.html
\bibitem[Bertin \& Arnouts(1996)]{Bertin1996} Bertin, E., \& Arnouts, S.\ 1996, \aaps, 117, 393
%\bibitem[Bertin(2006)]{Bertin2006} Bertin, E.\ 2006, Astronomical Data Analysis Software and Systems XV, 351, 112
\bibitem[Bowell et al.(1989)]{Bowell1989} Bowell, E., Hapke, B., Domingue, D., et al.\ 1989, Asteroids II, 524
%\bibitem[Chiorny et al.(2007)]{Chiorny2007} Chiorny, V.~G., Shevchenko, V.~G., Krugly, Y.~N., Velichko, F.~P., \& Gaftonyuk, N.~M.\ 2007, \planss, 55, 986
\bibitem[Donnison \& Wiper(1999)]{Donnison1999} Donnison, J.~R., \& Wiper, M.~P.\ 1999, \mnras, 302, 75
\bibitem[Fulchignoni et al.(1995)]{Fulchignoni1995} Fulchignoni, M., Barucci, M.~A., di Martino, M., \& Dotto, E.\ 1995, \aap, 299, 929
\bibitem[Grav et al.(2011)]{Grav2011} Grav, T., Mainzer, A.~K., Bauer, J., et al.\ 2011, \apj, 742, 40
\bibitem[Grillmair et al.(2010)]{Grillmair2010} Grillmair, C.~J., Laher, R., Surace, J., et al.\ 2010, Astronomical Data Analysis Software and Systems XIX, 434, 28
\bibitem[Harris(1996)]{Harris1996} Harris, A.~W.\ 1996, Lunar and Planetary Institute Science Conference Abstracts, 27, 493
\bibitem[Harris et al.(2012)]{Harris2012} Harris, A.~W., Pravec, P., \& Warner, B.~D.\ 2012, \icarus, 221, 226
%\bibitem[Harris \& Lagerros(2002)]{Harris2002} Harris, A.~W., \& Lagerros, J.~S.~V.\ 2002, Asteroids III, 205
\bibitem[Hergenrother \& Whiteley(2011)]{Hergenrother2011} Hergenrother, C.~W., \& Whiteley, R.~J.\ 2011, \icarus, 214, 194
\bibitem[Holsapple(2007)]{Holsapple2007} Holsapple, K.~A.\ 2007, \icarus, 187, 500
\bibitem[Jewitt et al.(2013)]{Jweitt2013} Jewitt, D., Ishiguro, M., \& Agarwal, J.\ 2013, \apjl, 764, L5
\bibitem[Law et al.(2009)]{Law2009} Law, N.~M., Kulkarni, S.~R., Dekany, R.~G., et al.\ 2009, \pasp, 121, 1395
\bibitem[Law et al.(2010)]{Law2010} Law, N.~M., Dekany, R.~G., Rahmer, G., et al.\ 2010, \procspie, 7735
\bibitem[Levitan et al.(2011)]{Levitan2011} Levitan, D., Fulton, B.~J., Groot, P.~J., et al.\ 2011, \apj, 739, 68
%\bibitem[Lomb(1976)]{Lomb1976} Lomb, N.~R.\ 1976, \apss, 39, 447
\bibitem[Mainzer et al.(2011)]{Mainzer2011} Mainzer, A., Grav, T., Bauer, J., et al.\ 2011, \apj, 743, 156
\bibitem[Masiero et al.(2009)]{Masiero2009} Masiero, J., Jedicke, R., {\v D}urech, J., et al.\ 2009, \icarus, 204, 145
\bibitem[Masiero et al.(2011)]{Masiero2011} Masiero, J.~R., Mainzer, A.~K., Grav, T., et al.\ 2011, \apj, 741, 68
\bibitem[Oey et al.(2007)]{Oey2007} Oey, J., Pray, D.~P., \& Pravec, P.\ 2007, Minor Planet Bulletin, 34, 101
\bibitem[Ofek et al.(2011)]{Ofek2011} Ofek, E.~O., Frail, D.~A., Breslauer, B., et al.\ 2011, \apj, 740, 65
\bibitem[Ofek et al.(2012a)]{Ofek2012a} Ofek, E.~O., Laher, R., Law, N., et al.\ 2012a, \pasp, 124, 62
\bibitem[Ofek et al.(2012b)]{Ofek2012b} Ofek, E.~O., Laher, R., Surace, J., et al.\ 2012b, \pasp, 124, 854
\bibitem[Parker et al.(2008)]{Parker2008} Parker, A., Ivezi{\'c}, {\v Z}., Juri{\'c}, M., et al.\ 2008, \icarus, 198, 138
\bibitem[Pilcher et al.(2009)]{Pilcher2009} Pilcher, F., Benishek, V., \& Krajewski, R.\ 2009, Minor Planet Bulletin, 36, 40
\bibitem[Polishook \& Brosch(2008)]{Polishook2008} Polishook, D., \& Brosch, N.\ 2008, \icarus, 194, 111
\bibitem[Polishook et al.(2012)]{Polishook2012} Polishook, D., Ofek, E.~O., Waszczak, A., et al.\ 2012, \mnras, 421, 2094
\bibitem[Pravec et al.(2006)]{Pravec2006} Pravec, P., Scheirich, P., Ku{\v s}nir{\'a}k, P., et al.\ 2006, \icarus, 181, 63
\bibitem[Pravec \& Harris(2000)]{Pravec2000} Pravec, P., \& Harris, A.~W.\ 2000, \icarus, 148, 12
\bibitem[Pravec et al.(2002)]{Pravec2002} Pravec, P., Ku{\v s}nir{\'a}k, P., {\v S}arounov{\'a}, L., et al.\ 2002, Asteroids, Comets, and Meteors: ACM 2002, 500, 743
\bibitem[Pravec et al.(2008)]{Pravec2008} Pravec, P., Harris, A.~W., Vokrouhlick{\'y}, D., et al.\ 2008, \icarus, 197, 497
\bibitem[Rau et al.(2009)]{Rau2009} Rau, A., Kulkarni, S.~R., Law, N.~M., et al.\ 2009, \pasp, 121, 1334
\bibitem[Rubincam(2000)]{Rubincam2000} Rubincam, D.~P.\ 2000, \icarus, 148, 2
\bibitem[Salo(1987)]{Salo1987} Salo, H.\ 1987, \icarus, 70, 37
%\bibitem[Scargle(1982)]{Scargle1982} Scargle, J.~D.\ 1982, \apj, 263, 835
\bibitem[Tedesco et al.(2005)]{Tedesco2005} Tedesco, E.~F., Cellino, A., \& Zappal{\'a}, V.\ 2005, \aj, 129, 2869
%\bibitem[van Eyken et al.(2011)]{VanEyken2011} van Eyken, J.~C., Ciardi, D.~R., Rebull, L.~M., et al.\ 2011, \aj, 142, 60
%\bibitem[Vokrouhlick{\'y} \& {\v C}apek(2002)]{Vokrouhlicky2002} Vokrouhlick{\'y}, D., \& {\v C}apek, D.\ 2002, \icarus, 159, 449
\bibitem[Warner et al.(2009)]{Warner2009} Warner, B.~D., Harris, A.~W., \& Pravec, P.\ 2009, \icarus, 202, 134
%\bibitem[Waszczak et al.(2013)]{Waszczak2013} Waszczak, A., Ofek, E.~O., Aharonson, O., et al.\ 2013, \mnras, 433, 3115
\bibitem[York et al.(2000)]{York2000} York, D.~G., Adelman, J., Anderson, J.~E., Jr., et al.\ 2000, \aj, 120, 1579

\end{thebibliography}
\end{document}